\def\Roman#1{\uppercase\expandafter{\romannumeral#1}}
\documentclass[a4paper,12pt]{article}
\usepackage{cite}
\usepackage{amssymb,amsfonts}
\usepackage[mathscr]{eucal}
\usepackage{mathrsfs}
\usepackage[utf8x]{inputenc}

\title{The Lagrange-Poincaré equations for a mechanical system with symmetry  on the principal fiber bundle 
over the base represented by the bundle space of the  associated bundle}

\author{S. N. Storchak\footnote{E-mail adress: storchak@ihep.ru}\\
\small{ Institute for High Energy Physics,
NRC ``Kurchatov Institute,''}\\
 \small{Protvino, 142284, Russian Federation}}

\begin{document}

  \maketitle

\begin{abstract}
The Lagrange--Poincar\'{e} equations for a mechanical system which describes the interaction of two scalar particles that move on a special Riemannian manifold, consisting of the product of two
manifolds, the total space of a principal fiber bundle and the vector space, are obtained.
The  derivation of equations is  performed by using the variational principle developed by Poincaré  for the mechanical systems with a symmetry. The obtained equations are written in terms of the
dependent variables which, as in gauge theories,  are implicitly determined by means of  equations representing the local sections  of the principal fiber bundle.
\end{abstract}

\section{Introduction}

Full research of mechanical systems suggests a finding of all critical points belonging to the systems, together with the behavior of phase curves in their vicinities. The knowledge of these points allows one to reconstruct the evolution of the mechanical systems.  At the equilibrium (the  point  which is  fixed in time under the dynamics), the evolution can be described, for instance, with the help of  the normal forms method proposed by   Poincar\'{e} and Birkhoff. 

In the case of dynamical systems with  symmetry, we are also interested in finding  the "fixed points". But  now they characterize the steady motions (or relative equilibria) of the original systems. These fixed points correspond to the equilibrium points of the reduced mechanical systems \cite{AbrMarsd,Marsden}.

The points of the relative equilibria are defined by the equations that follow from the Lagrange-Poincar{\'e} equations.
These last equations are represented by a system consisting of two equations. Usually they are referred to as  horizontal and vertical equations.\footnote{These equations are also  known as  Wong's equations \cite{Wong,Marsden_2}.}

Notice that such a description of the evolution is a consequence of the  choice of the special coordinate basis for the total space  of the  principal fiber bundle which can be associated with  the original mechanical  system with symmetry.

At present, the Lagrange-Poincar\'{e} equations  were obtained for many mechanical systems (see e.g. \cite{Marsden_1,Cendra} and references therein). 
The cited papers are dealing mainly with the systems that are invariant under the free and proper action of a Lie groups. The case of a non-free action was also considered. We refer to \cite{Montaldi}  were the necessary references may also be found.

The finite-dimensional mechanical systems with symmetry  are of interest for us not only by themselves.   Thanks to their properties, they can be also used as model systems in 
 studies that are carried out in gauge theories.

For example, an intrinsic geometry of the mechanical system, which describes the motion of the scalar particle on a Riemannian  manifold with a  free and proper  action of a group Lie, is similar to the geometry  of the gauge theory for  pure Yang--Mills fields \cite{Kunstatter}.

Due to the symmetry, the original motion of this particle can be viewed as occurring on the total space of the corresponding principal fiber bundle. In addition, as in gauge theories, the reduction process leads  to the ``true motion'' given on the reduced space -- the base space of the principal fiber bundle\cite{AbrMarsd}.

But in order to be able to use the mechanical systems with symmetry as a model systems for the gauge theories, it is necessary to have an appropriate  description   of the evolution of these mechanical systems.

In gauge theories, one usually deals with constrained variables, i.e. with the variables that are not independent but  satisfy  some functional equations.
These equations define (locally) the gauge surface in the space of gauge fields.
The gauge surface, in turn, determines the section of the corresponding trivial principal bundle. Hence the gauge surface is used for the coordinatization of the bundle space.
This means that in gauge theories, in the case of ``unresolved gauges'', that is when we are not able to find explicit solutions of the equations defining the gauge surface,   we have to use  dependent variables\footnote{They are the solutions of these equations.} (or locally, the dependent coordinates) for description of the evolution.
 It is these variables must be used in the Lagrange-Poincar\'{e} equations for model mechanical systems with symmetry, so that it would be possible to consider obtained equations as appropriate analogues for the corresponding equations of the gauge field theories. 

 This approach has been employed in  our works \cite{Storchak_wong_eq,Storchak_eq_Poinc}, where we have derived the Lagrange-Poincar\'{e} equations  for the above discussed  mechanical system with a symmetry. 
This enabled us to obtain  the equations for  Yang--Mills fields  having the same structure as the equations for the mechanical system. 
 As follows  from the Lagrange-Poincar\'{e} equations for   Yang--Mills fields \cite{Storchak_rel_equilibr_ym},  the defining equations for the relative equilibria are based on   the special spectral problem and therefore  may have  an  infinite number of solutions representing the possible relative equilibria.

We note that in the previously mentioned works  on the mechanical systems \cite{Marsden_1,Cendra, Montaldi}, as well as in  \cite{Ratiu}, where the Lagrange-Poincar\'{e} equations were obtained for the field theories, 
the questions related to the description of the evolution in terms of dependent variables  
have been left untouched.

Next   an important task in the approach based on using dependent variables, is to  determine the relative equilibria in the system consisting of the gauge field which interacts with a scalar field.
Before proceeding to its solution, and as a first step, it would be useful to examine an appropriate dynamical system in mechanics.
This is also necessary for verifying the correctness  of the corresponding equations in gauge theory.

In the present paper we consider a mechanical  system which describes a motion of two interacting scalar particles on a special configuration space -- the  product of two spaces. The first space in  this product is a finite-dimensional Riemannian manifold (without boundary), 
the second space  is a finite-dimensional vector space.
Moreover, we assume that there is a free, proper, isometric and smooth action of the compact Lie group  on these spaces and, therefore, on the original manifold as a whole.  It can be shown that such an action on  the first space, given by the  Riemannian manifold, leads to the principal fiber bundle in which  this Riemannian manifold is a total space.

As a result of a group action on the whole space, we also come  to the principal fiber bundle. For this bundle, the manifold, representing the original configuration space of  the whole system,  can be regarded  as a total space.
 But  the base of the bundle, i.e. the orbit space of the group action, now  is    the bundle   space of the associated vector bundle.

 We see, that  the obtained principal bundle is exactly the same which is used in the construction leading to  the associated vector bundle in geometry.
So the coordinates in the principal bundle are usually introduced by taking into account  this fact.

The coordinates in this principal bundle are determined  in  a standard way,  i.e. with the help of the local sections. 
The method, which we  use for the coordinatization of this principal bundle, is typical   for Yang--Mills fields with interactions.
In \cite{Huffel-Kelnhofer}, for example, it was  used in the study of the quantization procedure  in the  scalar electrodynamics.
In this method,  an important role belongs to  the  sections of the principal fiber bundle related to the Riemannian manifold representing  the first space in our configuration space.

The purpose of the present  paper is to obtain  the Lagrange-Poincar\'{e} equation for the model mechanical system which has been   described above. It will be made, as in \cite{Storchak_eq_Poinc}, with the help of the  Poincaré variational principle \cite{Poincare,Chetayev,Arnold}. This variational principle was developed by Poincar\'{e} for mechanical systems with a symmetry.

The paper will be organized as follows. In Section 2 we will introduce the principal fiber bundle coordinates on our original Riemannian manifold and  get a new representation for the metric tensor of this manifold.
In Section 3 we will change the coordinate basis of our manifold for the horizontal lift basis and also consider the corresponding transformation of the Lagrangian arising from it.
 In Section 4, after brief recalling the   Poincar\'{e} approach to the calculus of variations, we will derive the differential relations between the variations and the quasi-velocities. These relations are necessary for the Poincar\'{e} method. In Section 5 we will obtain the Lagrange-Poincar\'{e} equations.
The first of two Appendices will be dedicated to the derivation of the differential relation  
between the variations and the quasi-velocities associated with the group variables.  The second one
will contain the properties of the projection operators and Killing relations for the horizontal metric. In the last section will be given concluding remarks.

\section{Principal fiber bundle coordinates on the  configuration space}
The configuration space of our mechanical system, describing the interaction of two scalar particles, is represented by the product manifold $\mathcal P \times V$, where $\mathcal P$ is a smooth finite-dimensional  Riemannian manifold (without the boundary) and $V$ is a finite-dimensional vector space.

Let $(Q^A,f^n)$, $A=1,\dots , N_P$ and $n=1,\dots ,N_V$ be the coordinates of a point $(p,v)$ given on  a chart $(\mathcal U,\varphi ^{\tilde A})$ 
of the original product manifold. We assume that  open sets $\mathcal U$ of the charts are choosen to be equal $({\mathcal U}_P\times {\mathcal U}_V)$, and the coordinate functions for these sets are given by $\varphi ^{\tilde A}=(\varphi ^A,\varphi ^n)$.\footnote{In the sequel,  the expressions having the capital indices with the tilde mark will  be treated in a similar way.}
So, we have $Q^A=\varphi ^A(p)$ and $f^n=\varphi ^n(v)$.

In these coordinates, the Riemannian metric of the manifold is written as follows:
\begin{equation}
 ds^2=G_{AB}(Q)dQ^AdQ^B+G_{mn}df^mdf^n,
\label{metr_orig}
\end{equation}
where the first term represents the Riemannian metric of the manifold $\mathcal P$ and the second one, with  matrix $G_{mn}$ consisting of some fixed constant  elements, is used  as the metric of the inner product in $V$. 
It is admitted that $G_{mn}$ may be  non diagonal.
In the paper we assume that
we are given a free, proper, smooth and  isometric  action of the compact Lie group $\mathcal G$ on the original  manifold. Also we assume that the group acts on the manifold  from the right: $(p,v)g=(pg,g^{-1}v)$. 
Being written in coordinates, this action is given as follows:
\[
 {\tilde Q}^A=F^A(Q,g),\;\;\;\;{\tilde f}^n=\bar D^n_m(g)f^m.
\]
Here $\bar D^n_m(g)\equiv D^n_m(g^{-1})$,
and by $D^n_m(g)$ we denote the matrix of  the finite dimensional representation of the group $\mathcal G$
acting on the vector space $V$.

For the right action of the group $\mathcal G$ on the point $p$ with the coordinates $Q$, we have
\[
 F(F(Q,g_1),g_2)=F(Q,\rm \hat {\Phi}(g_1,g_2)),
\]
where the function $\rm\hat {\Phi}$  determines the group multiplication law in the space of the group parameters.

 Note that since the group $\mathcal G$ acts on $\mathcal P$ isometrically, the metric tensor $G_{AB}$ must satisfy  the following relation:
\begin{equation}
 G_{AB}(Q)=G_{DC}(F(Q,g))F^D_A(Q,g)F^C_B(Q,g),
\label{relat_G_AB}
\end{equation}
with $F^B_A(Q,g)\equiv\frac{\partial F^B(Q,g))}{\partial Q^A}$.  A similar relation for the tensor $G_{mn}$:
\begin{equation}
 G_{pq}=G_{mn}\bar D^m_p(g)\bar D^n_q(g),
\label{relat_g_mn}
\end{equation}
can be derived from the linear isometrical action of the group $\mathcal G$ on the vector space $V$.

The Killing vector fields for the product metric of the original manifold are  given by the corresponding vector fields for the manifolds $\mathcal P$ and $V$: 
$K^A_{\alpha}(Q)\frac{\partial}{\partial Q^A}$ with
$K^A_{\alpha}(Q)=\frac{\partial {\tilde Q}^A}{\partial a^{\alpha}}\Big|_{a=e}$ and
$K^n_{\alpha}(f)\frac{\partial}{\partial f^n}$ with $K^n_{\alpha}(f)=\frac{\partial {\tilde f}^n}{\partial a^{\alpha}}\Big|_{a=e}=\frac{\partial {\bar D}^n_m(a)}{\partial a^{\alpha}}\Big|_{a=e}=({\bar J}_{\alpha})^n_m f^m$.  The generators ${\bar J}_{\alpha}$ of the representation ${\bar D}^n_m(a)$ satisfy the  commutation relation 
$[{\bar J}_{\alpha},{\bar J}_{\beta}]={\bar c}^{\gamma}_{\alpha \beta}{\bar J}_{\gamma}$, where the structure constants
${\bar c}^{\gamma}_{\alpha \beta}=-{c}^{\gamma}_{\alpha \beta}$.

Using  the condensed notation, in which
$\tilde A\equiv (A,p)$, we can rewrite the components of the Killing vector fields as 
$$ K^{\tilde A}_{\mu}=(K^{A}_{\mu},K^{p}_{\mu}).$$

 We know from the general theory\cite{AbrMarsd} that the  action of a group  $\mathcal G$, provided that this action satisfies the same requirements as we  have  assumed in our paper,  leads to the orbit-fibering of the original manifold $\mathcal P\times V$. 
Therefore this manifold  can be regarded as the total space of the principal fiber bundle where the orbit space manifold $\mathcal P\times _{\mathcal G}V$ is the base space.
This means that it is possible to introduce  the principal fiber bundle coordinates in each local neighborhood of the original manifold and then to express the coordinates $(Q^A, f^n)$ of the point $(p,v)$ in terms of the  bundle coordinates. We briefly recall this well-known procedure 
\cite{Creutz, Razumov, Storchak_11, Storchak_12,Storchak_2,Storchak_3, Huffel-Kelnhofer}.

First we note that the  action of a group   $\mathcal G$ on $\mathcal P$  results in the principal fiber bundle $\rm P(\mathcal M,\mathcal G)$ having the base space  $\mathcal M= \mathcal P/\mathcal G$. 
The  coordinates of the points on  this bundle will be used as the part of  new coordinates on $\mathcal P\times V$.
 
The total space $\mathcal P$ of the principal fiber bundle $\rm P(\mathcal M,\mathcal G)$ has the following local representation $\pi ^{-1}(\mathcal U_x)\sim \mathcal U_x\times \mathcal G$, where $\mathcal U_x$ is an open neighbourhood of the point $x=\pi(p)$ which belongs to the chart $(\mathcal U_x,\varphi _x)$ of this bundle. It follows that the principal fiber bundle coordinates of the point $p$ are usually given by the coordinates $(x^i,a^{\alpha})$, $i=1,\dots,N_{\mathcal M},N_{\mathcal M}=\rm dim\, \mathcal M$, $\alpha =1,\dots,N_{\mathcal G},N_{\mathcal G}=\rm dim \,\mathcal G$ ($N_{\mathcal P}=N_{\mathcal M}+N_{\mathcal G}$).

As the coordinates on $\rm P(\mathcal M,\mathcal G)$, we    take in the paper not the coordinates $(x^i,a^{\alpha})$, since often it is difficult to find the interrelation between the new coordinates $x^i$ and the initial coordinates $Q^A$ of the point $p\in \mathcal P$, but  the constrained (or dependent)  coordinates. In gauge theories, such coordinates are determined  with the help of the gauge constraints and known also as adapted coordinates \cite{Huffel-Kelnhofer}. 
The same approach can be used in our case.

We suppose that in each sufficiently small neighbourhood of a point $p\in \mathcal P$ it is possible to determine   such a local submanifold $\Sigma$ of the manifold $\mathcal P$, which has a transversal intersection with  the orbits.
This property enables the existence of the local sections of the principal fiber bundle\cite{Bredon}. As a rule the submanifold $\Sigma$ is given by the equations $\{\chi^{\alpha}(Q)=0, \alpha= 1,\ldots N_{\mathcal G}\}$.   In gauge theories, the corresponding (functional)  equations  are known as  the gauge constraints. Hence, the submanifold $\Sigma$, which defines the local section of the principal fiber bundle, plays the same role as the  gauge fixing surface.  
 
The  coordinates  of the points on the local submanifold $\Sigma$ 
will be denoted by $Q^{\ast}{}^A$. Since they satisfy the equations $\{\chi^{\alpha}(Q^{\ast})=0\}$, they are called the dependent (or constrained) coordinates.   We will use these  coordinates, together with the group coordinates $a^{\alpha}$,  as the principal bundle coordinates for the points given on the total space $\mathcal P$ of the principal fiber bundle $\rm P(\mathcal M,\mathcal G)$.
 
The constrained coordinates  were exploited  in many works devoted to the  dynamical systems with  a symmetry (see e.g. \cite{Creutz, Sundermeyer, Falck, Razumov, Gavedzki,Kelnhofer} and  others).
The use of  these  coordinates for the coordinatization of the principal fiber bundles was considered in \cite{Babelon-Viallet, Mitter-Viallet,Kelnhofer}.  
 We will mainly follow   these works
where the   explanation of this procedure was given. Although  as the objects of the study of these works were chosen the gauge field theories, the approach, developed there, can  be also applied to our case. But, of course, it can be done only after the proper  adaptation of this approach to the mechanical systems.

As it is now well-known,
the introduction of the  coordinates in the principal fiber bundles is based on two statements. The first statement is concerned with the existence of the special locally finite open covering $\{U_{i}\}$ of the  base manifold $\mathcal M$, which is necessary for coordinate definition of the principal fiber bundle. It is assumed that this open covering is constructed with the help of the   set $\{\Sigma_{i}\}$ which is  formed by  the given local submanifolds (surfaces) $\Sigma_{i}$  of the total space $\mathcal P$.
Moreover, it is required that
a family local sections $\{\sigma _i\}$ of the principal fiber bundle $\pi :\mathcal P\to \mathcal M$ can be  determined by these local surfaces $\Sigma_{i}$: the section $\sigma_i$ is  the map $\sigma_i:U_i\to \Sigma_i$ such that  $\pi_{\Sigma_i}\cdot\sigma_i={\rm id}_{U_i}$.

The second statement is related to the definition  of the coordinate functions of the bundle charts  by which the coordinate principal fiber bundle is given. In a standard case, these functions are defined  by means of the rule which establishes the  local isomorphism between $U_{i}\times \mathcal G$ and $\pi ^{-1}(U_{i})$.  In \cite{Mitter-Viallet} and  \cite{Razumov}, the coordinate functions were obtained with the help of the local sections determined by means of parametrically given local surfaces $\Sigma_{i}$. In these works it was supposed  that the equations $\chi^{\alpha}(Q)=0$ had the following solution: $Q^A=Q^{\ast}{}^A(x)$, $x\in \mathcal M$.

But in many cases such a representation for the solution is impossible. This is the reason of using the dependent coordinates. So, to define  the coordinate functions of the principal fiber bundle arising in these cases one should  use the existing local isomorphism between trivial principal bundle $\Sigma_{i}\times \mathcal G\to\Sigma_{i}$ and 
$\rm P(\mathcal M,\mathcal G)$\cite{Babelon-Viallet,Huffel-Kelnhofer}. And now the coordinate functions of a bundle chart $(U_{i},\varphi_{i})$  perform  the following isomorphism:
\[
 \varphi_{i}:\,\,\Sigma_{i}\times \mathcal G\to \pi ^{-1}(U_{i}).
\]
In coordinates, this map is written as
\[
 \varphi_{i}:(Q^{\ast}{}^B,a^{\alpha})\to Q^A=F^A(Q^{\ast}{}^B, a^{\alpha}),
\]
where $Q^{\ast}{}^B$ are the coordinates of a  point given  on the local surface $ \Sigma_{i}$ and 
$a^{\alpha}$ -- the coordinates of an arbitrary group element $a$. This element carries the point, taken on $ \Sigma_{i}$, to  the point $p\in \mathcal P$ which has  the coordinates $Q^A$.

The inverse map $\varphi_{i}^{-1}$,
\[
 \varphi_{i}^{-1}:\,\,\pi ^{-1}(U_{i})\to\Sigma_{i}\times \mathcal G,
\]
has the following coordinate representation:
\[
 \varphi_{i}^{-1}: Q^A\to (Q^{\ast}{}^B(Q),a^{\alpha}(Q)).
\]
Here the group coordinates $a^{\alpha}(Q)$ of a point $p$ are the coordinates of the group element  
which connects, by means of its  action on $p$, the surface $\Sigma _{i}$ and the point $p\in \mathcal P$. These group coordinates are given by the solutions of the following equation:
\begin{equation}
 \chi^{\beta}(F^A(Q, a^{-1}(Q)))=0.
\label{a_chi}
\end{equation} 
The coodinates $Q^{\ast}{}^B$ are defined by the equation
\begin{equation}
 Q^{\ast}{}^B=F^B(Q, a^{-1}(Q)).
\label{Q_star}
\end{equation}
We see that the  map $\varphi_{i}^{-1}$, thus defined, enables one to find  the principal bundle coordinates $(Q^{\ast}{}^B,a^{\alpha})$ of the point $p$ from the known initial coordinates $Q^A$ of this point given on the manifold $\mathcal P$.

We note that the  bundle coordinates 
of $p\in\mathcal P$ were determined   for the  bundle chart $(U_{i},\varphi_{i})$ related to the local surface  $\Sigma_{i}$.
The relationship of these coordinates of the point $p$ with the coordinates obtained for the chart $(U_{j},\varphi_{j})$  is given by the transition function $\varphi_{ji}=\varphi ^{-1}_j\varphi_i$\cite{Huffel-Kelnhofer}:   
\[
 \varphi_{ji}:(\Sigma_{i}\cap\pi ^{-1}(U_j))\times\mathcal G\to  (\Sigma_{j}\cap\pi ^{-1}(U_i))\times\mathcal G.                
\]
In coordinates, this map is written as 
\[
 \varphi_{ji}: (Q^{\ast},a) \to \Bigl(F(Q^{\ast},a^{-1}_j(Q^{\ast})),{\hat \Phi}(a_j(Q^{\ast}),a)\Bigr),
  \]
where by $Q^{\ast}$ we denote the coordinates of the point belonging to $\Sigma_{i}$, $a_j(Q^{\ast})$ -- the coordinates of the group element defined by means of the local surface $\Sigma _j$ and by $a$ was denoted the coordinates of an arbitrary group element.

It is not difficult to check that these transition functions satisfy the cocycle relation
$\varphi _{ji}\varphi _{ik}=\varphi _{jk}$. It can be done by using the following formulae of the coordinate ransformations:
\[
g^{\alpha}(F(Q,a))=\hat \Phi^{\alpha}(g(Q),a),\;\;\;g^{-1}(F(Q,a))=\hat \Phi(a^{-1},g^{-1}(Q)),
\]
together with the general formulae for a group action:
$F(F(Q,g_1),g_2)=F(Q,\hat \Phi(g_1,g_2))$ and $\hat \Phi(a,\hat \Phi(g,h))=\hat \Phi(\hat \Phi(a,g),h)$.

For the principal fiber bundle ${\rm P}(\mathcal P\times _{\mathcal G}V,\mathcal G)$, 
adapted coordinates can be defined by the same method as it was done for $\rm P(\mathcal M,\mathcal G)$ \cite{Huffel-Kelnhofer}. Now $\pi: \mathcal P\times V\to \mathcal P\times _{\mathcal G}V$ means that locally we have a map $\pi: (p,v)\to [p,v]$, where  $[p,v]$ is the equivalence class formed by the equivalence relation   $(p,v)\sim (pg,g^{-1}v)$.  The local section $\tilde \sigma_i$ of this bundle, $\pi \cdot\tilde \sigma_i = \rm{id}$, is the map which sends $[p,v]$ to some element $(\tilde p,\tilde v)\in \mathcal P\times V$. The section $\tilde\sigma_i$ is given by
\[
 \tilde \sigma_i([p,v])=(\sigma_i(x),a(p) v), 
\]
where $\sigma_i$ is a local  section of $\rm P(\mathcal M,\mathcal G)$, $\sigma_i:U_i\to\pi_{\rm P}^{-1}(U_i)$, $x=\pi_{\rm P}(p)$ and $a(p)$ is the group element defined by $p=\sigma_i(x)a(p)$.
Since
\[
 (\sigma_i(x),a(p)\, v)=(p\,a^{-1}(p),a(p)\, v)=(p,v)\,a^{-1}(p),
\]
we get
\[
 \tilde \sigma_i([p,v])=(p,v)\,a^{-1}(p).
\]
We see that the image of $\tilde \sigma_i$, the local surface $\tilde \Sigma_i$, consists of the elements that are obtained in a similar way as the elements of ``gauge fixing surface'' $\Sigma_i \in\mathcal P$  
for the principal fiber bundle $\rm P(\mathcal M,\mathcal G)$.

For a properly chosen family of  sections $\{\tilde\sigma_i\}$, and, respectinely, the family of $\{\tilde \Sigma_i\}$,  the local isomorphisms of the principal fiber bundle ${\rm P}(\mathcal P\times _{\mathcal G}V,\mathcal G)$ and the  trivial principal bundles $\tilde \Sigma_i\times \mathcal G \to \tilde \Sigma_i$
enables one to introduce a new atlas on ${\rm P}(\mathcal P\times _{\mathcal G}V,\mathcal G)$  with charts that are related to the submanifolds $\{\tilde \Sigma_i\}$.

The coordinate functions  of these charts $(\tilde U_i,\tilde\varphi_i)$, where $\tilde U_i$ is an open neighborhood of the point $[p,v]$ given on the base space $\mathcal P\times_{\mathcal G}V$, are such that
\[
 \tilde \varphi_i^{-1} : \pi^{-1}(\tilde U_i)\to\tilde \Sigma_i \times \mathcal G,\;\;{\rm or}\; {\rm  in}\; {\rm coordinates,} 
\]
\[
 \tilde \varphi_i^{-1} :(Q^A,f^m)\to (Q^{\ast}{}^A(Q),\tilde f^n(Q),a^{\alpha}(Q)\,).
\]
Here  $Q^A$ and $f^m$ are the coordinates of a point $(p,v)\in \mathcal P\times V$,  
$Q^{\ast}{}^A(Q)$ is given by (\ref{Q_star}) and
\[
\tilde f^n(Q) = D^n_m(a(Q))\,f^m,
\]
$a(Q)$ is defined by  (\ref{a_chi}), and  we have used the following property: $\bar D^n_m(a^{-1})\equiv D^n_m(a)$.  The coordinates $Q^{\ast}{}^A$, representing  a point given on a local surface $\Sigma_i$,  satisfy the constraints: $\chi(Q^{\ast})=0$. That is, they are dependent coordinates.

The coordinate function $\tilde \varphi_i$ maps $\tilde \Sigma_i\times \mathcal G\to \pi^{-1}(\tilde U_i)$:
\[
 \tilde \varphi_i :(Q^{\ast}{}^B,\tilde f^n,a^{\alpha})\to (F^A(Q^{\ast},a), \bar D^m_n(a)\tilde f^n).
\]
Thus, we have defined the special  local bundle coordinates $(Q^{\ast}{}^A,\tilde f^n, a^{\alpha})$, also named as  adapted coordinates, 
on the principal fiber bundle 
$\pi:\mathcal P\times V\to \mathcal P\times_{\mathcal G} V$.

In the sequel we will deal, in fact, only  with the local expressions that are  given on a  separate chart. This case may be proper regarded, and also treated, 
by supposing that the principal fiber bundle $\rm P(\mathcal M, \mathcal G)$ is trivial.  It takes place, for example, when the local submanifolds $\{\chi ^{\alpha}=0\}$ form the global submanifold of the  manifold $\mathcal P$. Note that  
 ${\rm P}(\mathcal P\times_{\mathcal G} V, \mathcal G)$ will be also trivial.
For simplicity of further consideration, it will be  assumed in the paper that such a  restriction, imposed on the considered principal fiber bundles, is fulfilled.

As a consequence, we come to  a local isomorphism of the trivial principal  fiber bundle $\rm P(\mathcal M,\mathcal G)$ and the trivial principal  bundle 
$\pi_{\Sigma}:\Sigma \times \mathcal G \to \Sigma$ 
\cite{Mitter-Viallet,Huffel-Kelnhofer}. Therefore,  the charts of  the total space $\mathcal P$ are expressed through the charts of the global submanifold $\Sigma$. 
And  constrained global variables, defined on $\Sigma$, can be used as the coordinate functions of these charts.
It folows that  in this case, for the trivial principal fiber bundle ${\rm P}(\mathcal P\times_{\mathcal G} V, \mathcal G)$, we  have a bundle  isomorphism $\tilde \varphi: \tilde \Sigma\times \mathcal G\to \mathcal P\times V$ which enables us to define the charts with adapted coordinates on this bundle.

It is not difficult to obtain the representation for the Riemannian metric  given on $\mathcal P\times V$ in terms of the principal bundle  coordinates $(Q^{\ast}{}^A,{\tilde f}^n, a^{\alpha})$ which we have just introduced on the principal fiber bundle. The replacement of the coordinates $(Q^A,f^m)$ of a point $(p,v)\in \mathcal P\times V$  for a new coordinates
\begin{equation}
Q^A=F^A(Q^{\ast}{}^B,a^{\alpha}),\;\;\;f^m=\bar D^m_n(a)\tilde f^n
\label{transf_coord}
\end{equation}
 leads to the following
transformation of the local coordinate vector fields: 
\begin{eqnarray}
\displaystyle 
&&\!\!\!\!\!\!\!\!\frac{\partial}{\partial f^n}=D^m_n(a)\frac{\partial}{\partial {\tilde f}^m},
\nonumber\\
&&\!\!\!\!\!\!\!\!\frac{\partial}{\partial Q^B}=\frac{\partial Q^{\ast}{}^A}{ \partial Q^B}\frac{\partial}{\partial Q^{\ast}{}^A}+\frac{\partial a^{\alpha}}{\partial Q^B}\frac{\partial}{\partial a^{\alpha}}+\frac{\partial {\tilde f}^n}{\partial Q^B}\frac{\partial}{\partial {\tilde f}^n}
\nonumber\\
&&\!\!\!\!\!\!\!\!\!\!\!=\check F^C_B\Biggl(N^A_C(Q^{\ast})\frac{\partial}{\partial Q^{\ast}{}^A}+{\chi}^{\mu}_C({\Phi}^{-1})^{\beta}_{\mu}\bar{v}^{\alpha}_{\beta}(a)\frac{\partial}{\partial a^{\alpha}}-{\chi}^{\mu}_C({\Phi}^{-1})^{\nu}_{\mu}(\bar J_{\nu})^m_p\tilde f^p\frac{\partial}{\partial {\tilde f}^m}\Biggr).\:
\label{vectfield}
\end{eqnarray}
Here $\check F^C_B\equiv F^C_B(F(Q^{\ast},a),a^{-1})$ is an inverse matrix to the matrix $F^A_B(Q^{\ast},a)$,
${\chi}^{\mu}_C\equiv \frac{\partial {\chi}^{\mu}(Q)}{\partial Q^C}|_{Q=Q^{\ast}}$, $({\Phi}^{-1})^{\beta}_{\mu}\equiv({\Phi}^{-1})^{\beta}_{\mu}(Q^{\ast})$ -- the matrix which is inverse to the Faddeev--Popov matrix:
\[
 ({\Phi})^{\beta}_{\mu}(Q)=K^A_{\mu}(Q)\frac{\partial {\chi}^{\beta}(Q)}{\partial Q^A},
\]
the matrix $\bar{v}^{\alpha}_{\beta}(a)$ is inverse of the matrix $\bar{u}^{\alpha}_{\beta}(a)$.\footnote{$\det \bar{u}^{\alpha}_{\beta}(a)$ is   the density of the right-invariant measure  given on the group $\mathcal G$.}

The operator $N^A_C$, defined  as
\[
 N^A_C(Q)=\delta^A_C-K^A_{\alpha}(Q)({\Phi}^{-1})^{\alpha}_{\mu}(Q){\chi}^{\mu}_C(Q), 
\]
 is the projection operator ($N^A_BN^B_C=N^A_C$) onto the subspace which is orthogonal to the Killing vector field $ K^A_{\alpha}(Q)\frac{\partial}{\partial Q^A}$. $N^A_C(Q^{\ast})$ is the restriction of $N^A_C(Q)$ to the submanifold $ \Sigma $:
\[
 N^A_C(Q^{\ast})\equiv N^A_C(F(Q^{\ast},e))\;\;\;N^A_C(Q^{\ast})=F^B_C(Q^{\ast},a)N^M_B(F(Q^{\ast},a))\check F_M^A(Q^{\ast},a)
\]
$e$ is the unity element of the group. 

We note also that  formula (\ref{vectfield}) is a generalization of  an analogous formula
from \cite{Storchak_11,Storchak_2}.

As an operator, the vector field $\frac{\partial}{\partial Q^{\ast}{}^A}$ is determined  by means of the following rule:
\[
 \frac{\partial}{\partial Q^{\ast}{}^A }\varphi(Q^{\ast})=(P _\bot)^D_A(Q^{\ast})\frac{\partial \varphi(Q)}{\partial Q^D}\Bigl|_{Q=Q^{\ast}}
\]
where the projection operator $(P_\bot)^A_B$ on the tangent plane to the submanifold $\Sigma$ is given by
\[
 (P_\bot)^A_B=\delta^A_B-\chi ^{\alpha}_{B}\,(\chi \chi ^{\top})^{-1}{}^{\beta}_{\alpha}\,(\chi ^{\top})^A_{\beta}. 
\]
In this formula, $(\chi ^{\top})^A_{\beta}$ is a transposed matrix to the matrix $\chi ^{\nu}_B$:
\[
 (\chi ^{\top})^A_{\mu}=G^{AB}{\gamma}_{\mu\nu}\chi ^{\nu}_B\;\;\;\;{\gamma}_{\mu\nu}=K^A_{\mu}G_{AB}K^B_{\nu}.
\]
Using the above explicit expression
 for the projection operators, it is easy to derive 
their multiplication properties:
\[
 (P_\bot)^A_BN^C_A=(P_\bot)^C_B,\;\;\;\;\;N^A_B(P_\bot)^C_A=N^C_B.
\]

In the new coordinate basis 
$\displaystyle(\partial/\partial Q{}^{\ast A},\partial/\partial \tilde f^m,\partial/\partial a^{\alpha})$
the metric (\ref{metr_orig}) of the original manifold $\mathcal P \times V$ can be rewritten as follows:
\begin{equation}
\displaystyle
{\tilde G}_{\cal A\cal B}(Q{}^{\ast},\tilde f,a)=
\left(
\begin{array}{ccc}
 G_{CD}(P_{\bot})^C_A (P_{\bot})^D_B & 0 & G_{CD}(P_{\bot})^C_AK^D_{\nu}\bar u^{\nu}_{\alpha}\\
 0 & G_{mn} & G_{mp}K^p_{\nu}\bar u^{\nu}_{\alpha}\\
G_{BC}K^C_{\mu}\bar u^{\mu}_{\beta} & G_{np}K^p_{\nu}\bar u^{\nu}_{\beta} & d_{\mu\nu}\bar u^{\mu}_{\alpha}\bar u^{\nu}_{\beta}\\
\end{array}
\right)
\label{metric2c}
\end{equation}
where $G_{CD}(Q{}^{\ast})\equiv G_{CD}(F(Q{}^{\ast},e))$:
\[
 G_{CD}(Q{}^{\ast})=F^M_C(Q{}^{\ast},a)F^N_D(Q{}^{\ast},a)G_{MN}(F(Q{}^{\ast},a)), 
\]
the projection operators $P_\bot$ and the components $K^A_{\mu}$ of the Killing vector fields  depend on $Q{}^{\ast}$, $\bar u^{\mu}_{\beta}=\bar u^{\mu}_{\beta}(a)$,  $K^p_{\nu}=K^p_{\nu}(\tilde f)$,  $d_{\mu\nu}(Q{}^{\ast},\tilde f)\bar u^{\mu}_{\alpha}(a)\bar u^{\nu}_{\beta}(a)$ is 
the metric on  $\mathcal G$--orbit through the point $(p,v)$. The components $d_{\mu\nu}$ of this metric are  given by
\begin{eqnarray*}
 d_{\mu\nu}(Q{}^{\ast},\tilde f)&=&K^A_{\mu}(Q{}^{\ast})G_{AB}(Q{}^{\ast})K^B_{\nu}(Q{}^{\ast})+K^m_{\mu}(\tilde f)G_{mn}K^n_{\nu}(\tilde f)
\nonumber\\
&\equiv&\gamma_{\mu \nu}(Q^{\ast})+\gamma'_{\mu \nu}(Q^{\ast}).
\end{eqnarray*}

Also we note that when we made the change of the coordinates in the differential $df$:
$$df=\bar D^n_m(a)d{\tilde f}^m+\frac{\partial \bar D^n_m(a)}{\partial a^{\mu}}\tilde f^m da^{\mu},$$
we have used the  following transformations:
\[
 \frac{\partial \bar D^n_m(a)}{\partial a^{\mu}}\tilde f^m=(\bar J_{\beta})^l_m \bar D^n_l(a){\bar u}^{\beta}_{\mu}(a)\tilde f^m=K^l_{\beta}(\tilde f)\bar D^n_l(a){\bar u}^{\beta}_{\mu}(a).
\]
The last equality is due to 
the identity $D^l_k(a)(\bar J_{\alpha})^k_p\bar D^p_n(a)={\rho}^{\beta}_{\alpha}(a)(\bar J_{\beta})^l_n$, in which ${\rho}^{\beta}_{\alpha}(a)=\bar u^{\beta}_{\gamma}(a)v^{\gamma}_{\alpha}(a)$ is the matrix of the adjoint representation of the group $\mathcal G$.

The pseudoinverse matrix ${\tilde G}^{\cal A\cal B}(Q{}^{\ast},\tilde f,a)$ to  matrix (\ref{metric2c}) is as follows:
\begin{equation}
\displaystyle
\left(
\begin{array}{ccc}
{G}^{EF}N_E^AN_F^B & -G^{EF}N^A_E{\Lambda}^{\nu}_FK^m_{\nu} &  G^{EF}N^A_E{\Lambda}^{\beta}_F\bar v^{\alpha}_{\beta}\\
-G^{EF}N^A_F{\Lambda}^{\nu}_EK^m_{\nu} & G^{mn}+G^{EF}{\Lambda}^{\nu}_E{\Lambda}^{\mu}_FK^m_{\nu}K^n_{\mu} & -G^{EF}{\Lambda}^{\nu}_E{\Lambda}^{\mu}_FK^m_{\nu}{\bar v}^{\alpha}_{\mu}\\
G^{EF}N^B_F{\Lambda}^{\beta}_E\bar v^{\alpha}_{\beta} & -G^{EF}{\Lambda}^{\beta}_E{\Lambda}^{\mu}_FK^m_{\mu}{\bar v}^{\alpha}_{\beta} &G^{EF}{\Lambda}^{\nu}_E{\Lambda}^{\mu}_F{\bar v}^{\alpha}_{\nu}{\bar v}^{\beta}_{\mu}  \\
\end{array}
\right).
\label{metric2b}
\end{equation}
Here  ${\Lambda}^{\nu}_E\equiv({\Phi}^{-1})^{\nu}_{\mu}(Q{}^{\ast}){\chi}^{\mu}_E(Q{}^{\ast})$.

The pseudoinversion of ${\tilde G}_{\cal A\cal B}$ means that
\[
\displaystyle
{\tilde G}^{\tilde{\cal A}\tilde{\cal D}}{\tilde G}_{\tilde{\cal D}\tilde{\cal B}}=
\left(
\begin{array}{ccc}
  (P_{\bot})^A_B & 0 & 0\\
 0 & {\delta}^a_b & 0\\
0 & 0 & {\delta}^{\alpha}_{\beta}\\
\end{array}
\right).
\]

\section{Transformation of the Lagrangian}

In terms of  initial  local coordinates  defined on the original manifold $\mathcal P\times V$, the  Lagrangian for the considered mechanical system can be written  as follows:
\begin{equation}  
\mathcal L=\frac12 G_{AB}(Q)\,{\dot Q}^A{\dot Q}^B +\frac12 G_{mn}\,{\dot f}^m{\dot f}^n-V(Q,f).
\label{lagrang_1}
\end{equation}
 By our assumption,  the potential $V(Q,f)$ is a $\mathcal G$-invariant function: $V(Q,f)=V(F(Q,a),\bar D(a)f)$. So the whole Lagrangian is also invariant.

As the configuration space $\mathcal P\times V$ of our mechanical system is a total space of the principal fiber bundle ${\rm P}(\mathcal P\times_{\mathcal G}V,\mathcal G)$, the evolution of the system may be equally represented by using the bundle coordinates. 
In particular, in the Lagrangian (\ref{lagrang_1}), new coordinates  are introduced by using
the replacement of the local coordinates  (\ref{transf_coord}). As a result, we get
\begin{eqnarray}
 &&{\mathcal L}=\frac12G_{CD}\Bigl(\frac{d{Q^{\ast}}^{C}}{dt}+K^C_{\mu}\,{\bar u}^{\mu}_{\alpha}(a)\,\frac{da^{\alpha}}{dt}\Bigr)\Bigl(\frac{d{Q^{\ast}}^{D}}{dt}+K^D_{\nu}\,{\bar u}^{\nu}_{\beta}(a)\,\frac{da^{\beta}}{dt}\Bigr)
\nonumber\\
&&\;\;\;\;\;+\frac12G_{mn}\Bigl(\frac{d{\tilde f^m}}{dt}+K^m_{\beta}\,{\bar u}^{\beta}_{\alpha}(a)\,\frac{da^{\alpha}}{dt}\Bigr)\Bigl(\frac{d{\tilde f^n}}{dt}+K^n_{\nu}\,{\bar u}^{\nu}_{\mu}(a)\,\frac{da^{\mu}}{dt}\Bigr)-V,
\label{lagrang_2}
\end{eqnarray}
where now $G_{CD}, K^C_{\mu}$ depend on $Q^{\ast}$, $K^m_{\beta}=K^m_{\beta}(\tilde f)$, and $V=V(Q^{\ast},\tilde f)$.

We note that transformation of the velocities $\dot Q^A(t)$ in (\ref{lagrang_1}), for 
$Q^A(t)=F^A(Q^{\ast}{}^D(t), a^{\alpha}(t))$, was made as follows:
\begin{eqnarray*}
{\dot Q}^A(t)&\equiv&\frac{dQ^A}{dt}=F^A_C\,({P_{\bot}})^C_D\,\frac{d{Q^{\ast}}^D}{dt}+F^A_{\alpha}\,\frac{da^{\alpha}}{dt}
\nonumber\\
&=&F^A_C\,  \Bigl(\frac{d{Q^{\ast}}^C}{dt}+K^C_{\beta}(Q^{\ast})\,{\bar u}^{\beta}_{\alpha}(a)\,\frac{da^{\alpha}}{dt}\Bigr),
\end{eqnarray*}
where $F^A_{\alpha}\equiv\frac{\partial F^A(Q^{\ast},a)}{\partial a^{\alpha}}=F^A_CK^C_{\beta}{\bar u}^{\beta}_{\alpha}$. Besides,  we have used 
the identity $({P_{\bot}})^C_D\,\frac{d{Q^{\ast}}^D}{dt}=\frac{d{Q^{\ast}}^C}{dt}$, which is valid since  the velocity $dQ^{\ast}{}^D\!/dt$ belongs to the tangent space ${\rm T}\,\Sigma$ of the gauge fixing surface $\Sigma=\{Q^{\ast}{}^A:\chi ^{\alpha}(Q^{\ast})=0\}$. 

Our next task is to introduce a special coordinate basis (the horizontal lift basis) on the total space of the principal fiber bundle. This basis 
 is  needed for derivation of the Lagrange-Poincar\'e  equations  in the considered problem. 
We note that  coordinate vector fields of this basis do not commute between themselves. Sometimes, mainly in  a physical literature, such bases are called the nonholonomic. In our works \cite{Storchak_wong_eq, Storchak_eq_Poinc}, an analogous basis was constructed for ${\rm P}(\mathcal M,\mathcal G)$.

Since the new basis consists of the horizonal and vertical vector fields, this means that there is a connection on the considered principal fiber bundle.
In  case of the reduction of  mechanical systems, it is this case we study here, there exists \cite{AbrMarsd} a natural connection called the ''mechanical connection``. So, it is quite natural that this connection  takes part in the process of the  separation of  vector fields into two
orthogonal sets.
We briefly recall how one can introduce these vector fields.

The one-form  $\hat\omega ^{\alpha}$   on the principal fiber bundle ${\rm P}(\mathcal P\times_{\mathcal G}V,\mathcal G)$,\footnote{The one-form $\hat\omega$ with the value in the Lie algebra of the group Lie $\mathcal G$ is  $\hat\omega=\hat\omega ^{\alpha}\otimes\lambda_{\alpha}$.} the connection form, is given by the following formula written in terms of the initial local coordinate given  
on the total space $\mathcal P\times V$:
\begin{equation}
 \hat\omega ^{\alpha}(Q,f)=d^{\alpha\beta}(Q,f)\,(K^B_{\beta}(Q)G_{BA}(Q)dQ^A+K^p_{\beta}(f)G_{pq}df^q\,).
\label{connect_Q}
\end{equation}
To rewrite this connection form in terms of the principal fiber bundle coordinates $(Q^{\ast}{}^A,\tilde f^n,a^{\alpha})$, one should perform the corresponding transformations  of all terms in this expression. It can be made as follows:
\[
 dQ^A=F^A_{A'}\,((P_{\bot})^{A'}_{D}dQ{}^{\ast D}+K^{A'}_{\mu}(Q{}^{\ast})\bar u^{\mu}_{\alpha}(a)da^{\alpha}),
\]
with  $(P_{\bot})^{A'}_{D}dQ{}^{\ast D}=dQ{}^{\ast { A'}}$,
\[
 df^q=\bar D^q_n(a)\,(d\tilde f^n+K^n_{\mu}(\tilde f)\bar u^{\mu}_{\alpha}(a)da^{\alpha}).
\]
 \[
 K^B_{\beta}(F(Q^{\ast},a))={\rho}^{\mu}_{\beta}(a)K^D_{\mu}(Q^{\ast})F^B_D(Q^{\ast},a),
\]
and
\[
 K^p_{\beta}(\bar D(a)\tilde f)={\rho}^{\mu}_{\beta}(a)K^q_{\mu}(\tilde f)\bar D^p_q(a).
\]
\[
d^{\alpha\beta}(Q,f)=\bar {\rho}^{\alpha}_{\alpha '}(a) \bar {\rho}^{\beta}_{\beta '}(a)d^{\alpha '\beta '}(Q{}^{\ast},\tilde f).
\]
To transform $G_{BA}(Q)$ and $G_{pq}$ one must takes into account the relations (\ref{relat_G_AB}) and (\ref{relat_g_mn}).
The above transformations leads to
\[
 \hat\omega ^{\alpha}=\bar {\rho}^{\alpha}_{\alpha '}(a)\biggl(d^{\alpha '\mu}K^D_{\mu}(Q{}^{\ast})G_{DA}(Q{}^{\ast})dQ{}^{\ast  A}+d^{\alpha '\mu}K^q_{\mu}(\tilde f)G_{qn}d{\tilde f}^n\biggr)+u^{\alpha}_{\beta}(a)da^{\alpha},
\]
where now  $d^{\alpha '\mu}= d^{\alpha '\mu}(Q{}^{\ast},\tilde f)$.
The obtained expression for $\hat\omega ^{\alpha}$ may be rewritten as follows:
\begin{equation}
\hat\omega ^{\alpha}= 
\tilde\mathscr A^{\alpha '}_B(Q{}^{\ast},\tilde f,a)dQ{}^{\ast  B}+\tilde\mathscr A^{\alpha '}_m(Q{}^{\ast},\tilde f,a)d{\tilde f}^m+u^{\alpha}_{\beta}(a)da^{\alpha},
\label{connect_Q_star}
\end{equation}
where we have introduced the (gauge) potentials $\mathscr A^{\alpha }_B$, and $\mathscr A^{\alpha '}_m$, together with  a new notation: $\tilde\mathscr A^{\alpha }_B=\bar {\rho}^{\alpha}_{\alpha '}(a)\mathscr A^{\alpha '}_B(Q{}^{\ast},\tilde f)$.

 The same may be written in the condensed notation:
\[
 \hat\omega ^{\alpha}= 
\tilde\mathscr A^{\alpha '}_{\tilde B}(Q{}^{\ast},\tilde f,a)dQ{}^{\ast  \tilde B}+u^{\alpha}_{\beta}(a)da^{\alpha},
\]
 implying that the index represented by  the capital Latin letter with the tilde mark has  two components: $\tilde B=(B,p)$ and, respectively, the variables are given as 
$Q{}^{\ast  \tilde B}=(Q{}^{\ast  B},\tilde f^p)$.   The condensed
notation  will be also used in the sequel,

We note that  analogous  transformations convert the Killing vector field $K_{\alpha}(Q,f)$, the vertical vector field,
\[
K_{\alpha}(Q,f)= K^B_{\alpha}(Q)\frac{\partial}{\partial Q^B}+K^p_{\alpha}\frac{\partial}{\partial { f}^p},
\]
into the vector field $L_{\alpha}=v^{\nu}_{\alpha}(a)\frac{\partial}{\partial a^{\nu}}$ which is   the left-invariant vector field  given on the group manifold $\mathcal G$.
 
In order to define the horizontal vector fields, we  need  the horizontal projection operators. These operators must extract the direction which is normal to the orbit: ${\Pi}^{\tilde A}_{\tilde E}K^{\tilde E}_{\alpha}=0$. They are defined as follows:
$${\Pi}^{\tilde A}_{\tilde B}={\delta}^{\tilde A}_{\tilde B}-K^{\tilde A}_{\alpha}d^{\alpha \beta}K^{\tilde D}_{\beta}G_{\tilde D \tilde B}.$$
By  ${\Pi}^{\tilde A}_{\tilde B}$, we denote the  four component operator:
$${\Pi}^{\tilde A}_{\tilde B}=({\Pi}^{A}_{B},\, {\Pi}^{A}_{m},\,{\Pi}^{m}_{A},\,{\Pi}^{m}_{n}).$$
The components  are given by the following formulae:
\begin{eqnarray*}
&&{\Pi}^{A}_{B}={\delta}^A_B-K^{A}_{\alpha}d^{\alpha \beta}K^{D}_{\beta}G_{D B}\\
&&{\Pi}^{A}_{m}=-K^{A}_{\mu}d^{\mu \nu}K^{p}_{\nu}G_{pm}\\
&&{\Pi}^{m}_{A}=-K^{m}_{\mu}d^{\mu \nu}K^{D}_{\nu}G_{D A}\\
&&{\Pi}^{m}_{n}={\delta}^m_n-K^{m}_{\mu}d^{\mu \nu}K^{r}_{\nu}G_{rn}.\\
\end{eqnarray*}

The horizontal vector fields are defined as follows:
\begin{equation}
 H_A(Q,f)={\Pi}^R_A\frac{\partial}{\partial Q^R}+{\Pi}^q_A\frac{\partial}{\partial f^q}
\label{H_A(Q)}
\end{equation}
\begin{equation}
 H_p(Q,f)={\Pi}^R_p\frac{\partial}{\partial Q^R}+{\Pi}^m_p\frac{\partial}{\partial f^m}
\label{H_p(Q)}.
\end{equation}
For the connection form $\hat\omega ^{\alpha}$ from (\ref{connect_Q}), one can easily check
the fulfilment of the following equalities: 
\[
 \hat\omega ^{\alpha}(H_A)=0,\;\;\;\hat\omega ^{\alpha}(H_p)=0,\;\;\;\hat\omega ^{\alpha}(K_{\beta})=\delta ^{\alpha}_{\beta}.
\]
This means that $H_A$  and $H_p$ are the horizontal vector fields. The Killing vector field $K_{\beta}$ is the vertical one.

With formulae (\ref{vectfield}) and the  properties of the projection operator 
${\Pi}^{\tilde A}_{\tilde B}$ that are given in  Appendix, we may express the horizontal vector fields (\ref{H_A(Q)}) and (\ref{H_p(Q)}) in terms of the principal fiber bundle coordinates. They are given as follows:
\[
 H_A(F^B(Q^{\ast}, a), \bar D^r_p(a)\tilde f^p)=\check F^M_A\,H_M(Q^{\ast},\tilde f,a),
\]
where
\begin{equation}
H_M(Q^{\ast},\tilde f,a)= \Bigl[N^T_M\Bigl(\frac{\partial}{\partial Q^{\ast T}}-\tilde \mathscr A^{\alpha }_T L_{\alpha}\Bigl)+N^m_M\Bigl(\frac{\partial}{\partial {\tilde f}^m}-\tilde \mathscr A^{\alpha }_mL_{\alpha}\Bigl)\Bigr],
\label{H_A(Q_star)}
\end{equation}
and 
\[
 H_p(F^B(Q^{\ast}, a), \bar D^r_p(a)\tilde f^p)=D^m_p(a)H_m(Q^{\ast},\tilde f,a),
\]
where
\begin{equation}
 H_m(Q^{\ast},\tilde f,a)=\Bigl( \frac{\partial}{\partial {\tilde f}^m}-\tilde \mathscr A^{\alpha }_mL_{\alpha}\Bigl)
\label{H_p(Q_star)}.
\end{equation}
In  equation (\ref{H_A(Q_star)}), we use  the components of the projection operator $N^{\tilde A}_{\tilde C}$ :
$$N^{\tilde A}_{\tilde C}=(N^A_C,N^A_m,N^m_A,N^m_p).$$ 
Besides of $N^A_C$, which was already defined, the components are 
\[
N^A_m=0,\;\;\;  N^m_A=-K^m_{\alpha}({\Phi}^{-1})^{\alpha}_{\mu}\,{\chi}^{\mu}_A=-K^m_{\alpha}{\Lambda}^{\alpha}_A,\;\;\;  N^m_p={\delta}^m_p.
\]
The operator $N^{\tilde A}_{\tilde B}$ satisfy the following properties:
\[
N^{\tilde A}_{\tilde B} N^{\tilde B}_{\tilde C}=N^{\tilde A}_{\tilde C},\;\;\;\;
{\Pi}^{\tilde L}_{\tilde B}N^{\tilde A}_{\tilde L}=N^{\tilde A}_{\tilde B},\;\;\;
{\Pi}_{\tilde L}^{\tilde A}N_{\tilde C}^{\tilde L}={\Pi}_{\tilde C}^{\tilde A}.
\]

The horizontal vector fields  that are defined by the formulae (\ref{H_A(Q_star)}) and (\ref{H_p(Q_star)}), together with  the left-invariant vector field $L_{\alpha}$,  represent the new local coordinate basis for our principal fiber bundle. 
The horizontal coordinate vector fields of this basis do not commute between themselves. They commutation relations are as follows:

\begin{equation}
 [H_A,H_B]={\mathbb C}^T_{AB}\,H_T+{\mathbb C}^p_{AB}\,H_p+{\mathbb C}^{\alpha}_{AB}L_{\alpha}, 
\label{commrelat_AB}
\end{equation}
where the ``structure constants'' are given by 

\[{\mathbb C}^T_{AB}=({\Lambda}^{\gamma}_A N^R_B-{\Lambda}^{\gamma}_BN^R_A) K^{T}_{{\gamma} R},\]

\[{\mathbb C}^p_{AB}=-N^D_AN^R_B({\Lambda}^{\alpha}_{R,D}-{\Lambda}^{\alpha}_{D,R}) K^p_{\alpha}  \;\;  -c^{\sigma}_{\alpha \beta}{\Lambda}^{\beta}_A{\Lambda}^{\alpha}_BK^p_{\sigma}, \]

and
\[{\mathbb C}^{\alpha}_{AB}=-N^S_AN^P_B\,\tilde{\mathcal F}^{\alpha}_{SP}-(N^E_AN^p_B-N^E_BN^p_A)\tilde{\mathcal F}^{\alpha}_{Ep}+N^m_AN^p_B\tilde{\mathcal F}^{\alpha}_{pm}\,.
\]
In ${\mathbb C}^T_{AB}$,  we denote the partial derivative of $K^{T}_{{\gamma}}$ with respect to $Q^{\star}{}^R$ by $K^{T}_{{\gamma} R}$. In ${\mathbb C}^{\alpha}_{AB}$,
the curvature tensor $\tilde{\mathcal F}^{\alpha}_{SP}$ of the connection ${\tilde{\mathscr A}^{\alpha}_P}$ is given by
\[
\tilde{\mathcal F}^{\alpha}_{SP}=\displaystyle\frac{\partial}{\partial Q^{\ast}{}^S}\,\tilde{\mathscr A}^{\alpha}_P- 
\frac{\partial}{\partial {Q^{\ast}}^P}\,\tilde{\mathscr A}^{\alpha}_S
+c^{\alpha}_{\nu\sigma}\, \tilde{\mathscr A}^{\nu}_S\,
\tilde{\mathscr A}^{\sigma}_P,
\]
($\tilde{\mathcal F}^{\alpha}_{SP}({Q^{\ast}},a)={\bar{\rho}}^{\alpha}_{\mu}(a)\,{\mathcal F}^{\mu}_{SP}(Q^{\ast})\,$).   The tensors $\tilde{\mathcal F}^{\alpha}_{Ep}$ and $\tilde{\mathcal F}^{\alpha}_{pm}$ are defined in a similar way.

Next commutation relations are 
\begin{equation}
 [H_A,H_p]={\mathbb C}^m_{Ap}\,H_m+{\mathbb C}^{\alpha}_{Ap}L_{\alpha}
\label{commrelat_Ap}
\end{equation}
with
\[
{\mathbb C}^m_{Ap}=({\bar J}_{\alpha})^m_p{\Lambda}^{\alpha}_A,\;\;\;
{\mathbb C}^{\alpha}_{Ap}=-N^E_A\tilde{\mathcal F}^{\alpha}_{Ep}-N^m_A\tilde{\mathcal F}^{\alpha}_{mp},
 \]
and
\begin{equation}
 [H_p,H_q]={\mathbb C}^{\alpha}_{pq}L_{\alpha}
\label{commrelat_pq}
\end{equation}
with
\[
 {\mathbb C}^{\alpha}_{pq}=-\tilde{\mathcal F}^{\alpha}_{pq}\,.
\]

We notice that the left-invariant vector fields $L_{\alpha}$ of the new basis commute with the coordinate  horizontal vector fields:
\[
 [H_A,L_{\alpha}]=0,\;\;\;[H_p,L_{\alpha}]=0.
\]
Also, for $L_{\alpha}$ we have $[L_{\alpha},L_{\beta}]=c^{\gamma}_{\alpha \beta}L_{\gamma}$.

In the new coordinate basis $(H_A,H_p,L_{\alpha})$,  the metric (\ref{metric2c}) can be  written as the metric 
$\check G_{\mathcal A \mathcal B}$ with following components:
\begin{equation}
\displaystyle
{\check G}_{\cal A\cal B}(Q^{\ast},\tilde f,a)=
\left(
\begin{array}{ccc}
{\tilde G}^{\rm H}_{AB} & {\tilde G}^{\rm H}_{Am} & 0\\
{\tilde G}^{\rm H}_{nB} & {\tilde G}^{\rm H}_{nm} & 0\\
0 & 0 &\tilde{d}_{\alpha \beta }  \\
\end{array}
\right)=\left( \begin{array}{cc}
{\tilde G}^{\rm H}_{\tilde A \tilde B}  & 0 \\
0 & \tilde{d}_{\alpha \beta }  \\
\end{array}
\right),
\label{metric2cc}
\end{equation}
where $\tilde{d}_{\alpha \beta }=\rho^{\alpha'}_{\alpha}\rho^{\beta'}_{\beta}d_{\alpha' \beta' }$. 
The  components of the ``horizontal metric'' ${\tilde G}^{\rm H}_{\tilde A \tilde B}$ depending on $(Q^{\ast}{}^A,\tilde f ^m)$ are defined as follows:
$${\tilde G}^{\rm H}_{AB}={\Pi}^{\tilde A}_{A}\,{\Pi}^{\tilde B}_B \,G_{\tilde A\tilde B}=G_{AB}-G_{AD}K^{D}_{\alpha}d^{\alpha \beta}K^R_{\beta}\,G_{RB},$$
because of ${\Pi}^{\tilde C}_{A}\,{\Pi}^{\tilde D}_B \,G_{\tilde C\tilde D}= {\Pi}^{C}_{A}\,{\Pi}^{D}_B \,G_{CD}+
{\Pi}^{q}_{A}\,{\Pi}^{p}_B \,G_{qp}$.

$${\tilde G}^{\rm H}_{Am}=-G_{AB}K^{B}_{\alpha}\,d^{\alpha \beta}K^p_{\beta}G_{pm}.$$
Notice that ${\tilde G}^{\rm H}_{Am}$ is equal to
$${\tilde G}^{\rm H}_{mA}=-G_{mq}K^{q}_{\mu}\,d^{\mu \nu}K^D_{\nu}G_{DA}.$$

$${\tilde G}^{\rm H}_{mn}={\Pi}^r_mG_{rn},\;\;\;\; \rm{or}$$
${\Pi}^{\tilde C}_{m}\,{\Pi}^{\tilde D}_n \,G_{\tilde C\tilde D}= {\Pi}^{C}_{m}\,{\Pi}^{D}_n \,G_{CD}+
{\Pi}^{r}_{m}\,{\Pi}^{q}_n \,G_{rq}=G_{mn}-G_{mr}K^r_{\alpha}d^{\alpha \beta}K_{\beta}^pG_{pn}.$

It worth to note that metric ${\tilde G}^{\rm H}_{\tilde\mathcal A \tilde\mathcal B}$  is given on the local surface $\tilde \Sigma$ and gives rise the metric on the orbit space $\mathcal P\times _\mathcal G V$ provided that the submanifold  $\tilde \Sigma$ is given parametrically.

The pseudoinverse matrix ${\check G}^{\cal A\cal B}$ to the matrix (\ref{metric2cc}) is represented as
\begin{equation}
\displaystyle
{\check G}^{\cal A\cal B}=
\left(
\begin{array}{ccc}
{G}^{EF}N^A_EN^B_F & {G}^{EF}N^A_EN^q_F & 0\\
{G}^{EF}N^p_F N^B_E & {G}^{pq}+G^{AB}N^p_AN^q_B & 0\\
0 & 0 &\tilde{d}^{\alpha \beta }  \\
\end{array}
\right).
\label{metric2bb}
\end{equation}
This matrix is defined from the following orthogonality condition:
\[
\displaystyle
{\check G}^{\cal A\cal B}{\check G}_{\cal B\cal E}=
\left(
\begin{array}{ccc}
N^A_D& 0 & 0\\
 N^p_D & {\delta}^p_m & 0\\
0 & 0 & {\delta}^{\alpha}_{\beta} \\
\end{array}
\right)=
\left(
\begin{array}{cc}
N^{\tilde A}_{\tilde D} & 0 \\
0 & {\delta}^{\alpha}_{\beta} \\
\end{array}
\right),
\]
where 
\[
\displaystyle
N^{\tilde A}_{\tilde D}=
\left(
\begin{array}{cc}
N^A_D& N^A_m\\
 N^p_D & N^p_m\\
\end{array}
\right)
\]
($N^A_m=0, N^p_m={\delta}^p_m$).

In the local coordinates of the  basis $(H_A,H_p,L_{\alpha})$, the expression (\ref{lagrang_2}) of the Lagrangian $\mathcal L$ is transformed into
\begin{equation}
 {\hat{\mathcal L}}=\frac12\,({\tilde G}^{\rm H}_{AB}\, {\omega}^A {\omega}^B +{\tilde G}^{\rm H}_{Ap}\, {\omega}^A {\omega}^p+{\tilde G}^{\rm H}_{pA}\, {\omega}^p {\omega}^A+{\tilde G}^{\rm H}_{pq}\, {\omega}^p {\omega}^q+ {\tilde{d}}_{\mu \nu} {\omega}^{\mu} {\omega}^{\nu})-V,
\label{lagrang_3}
\end{equation}
where we have introduced the new time-dependent variables ${\omega}^A,{\omega}^p$ and ${\omega}^{\alpha}$ that are related to the velocities:
\begin{eqnarray} 
&&\omega ^A=(P_{\bot})^A_B\, \frac{dQ^{\ast}{}^B}{dt}=\frac{dQ^{\ast}{}^A}{dt},\;\;\;\;
\omega ^p=\frac{d\tilde f^p}{dt}\nonumber\\
&&{\omega}^{\alpha}=u^{\alpha}_{\mu}\frac{da^{\mu}}{dt}+{\tilde {\mathscr A}}^{\alpha}_E\,\frac{dQ^{\ast}{}^E}{dt}+{\tilde {\mathscr A}}^{\alpha}_m\, \frac{d\tilde f^m}{dt}.
\label{veloc_omega}
\end{eqnarray}

\section{Differential relations between variations and quasi-velocities}

The main peculiarity of the Poincar\'{e}  variational principle 
is that it exploits the special variations. These  variations are connected with  the  independent vector fields,  which, unlike of the usual case,   may not commute between themselves.
If these vector fields $v_1,\ldots,v_n$,  given on a  some smooth manifold, form a basis, then, in general, their commutator is as follows: $[v_i,v_j]=c^k_{ij}(q)v_k$, where the ``structure constants'' are represented by the functions  on this manifold.

For a smooth path  $q^i(t)$, defined  on a  manifold, the time derivative of the function $f$ taken on this path  is determined  as 
\begin{equation}
 \frac{df(q(t))}{dt}=\frac{\partial f}{\partial q^i} \frac{d q^i}{d t}={\sum}_iv_i(f) {\omega}^i,
\label{f_dt}
\end{equation}
where $v_i(f)$ is the directional derivative of $f$ along the vector field $v_i$.
The variables ${\omega}^i$ are called  the quasi-velocities. They are  linear functions of the velocities $\dot q^i(t)$.

In the approach of Poincar\'{e}, 
the deformation of the path $q(t)$, which we will   denote  by $q(u,t)$, such that $q(0,t)=q(t)$, 
have the standard properties. We refer to \cite{Arnold}, where for the variations with the fixed ends, these properties were considered.

But as for the derivative of the function $f(q(u,t))$ with respect to the deformation parameter $u$,  it is given by the following  expression:
 \begin{equation}
 \frac{\partial f(q(u,t))}{\partial u}={\sum}_iv_i(f) {w^i}(u,t).
\label{f_du}
\end{equation}
 Also, it is required that the  variations $w^i(u,t)$  are independent within  the time interval $[t_1,t_2]$ of the considered variational problem.  And at the ends of the time interval, they satisfy the following conditions: $w^k(u,t_1)=0$ and $w^k(u,t_2)=0$.

The variation of the functional $F(q(t))$  in this variational calculus is obtained by the usual method:
\begin{equation}
 \delta F =\frac{dF(q(u,t))}{du}\Bigl|_{u=0}.
\label{var_funk}
\end{equation} 

We will apply  the Poincar\'{e} variational principle  to the action functional
\begin{equation}
 S=\int_{t_1}^{t_2}{\hat{\mathcal L}}\,dt,
\label{action} 
\end{equation}
with the Lagrangian (\ref{lagrang_3}).

But first of all, we have to obtain the  differential relations between  the quasi-velocities 
and the variations. 
These relations are necessary in order to calculate the  variations  of the functionals by the Poincar\'{e} method.

In our case we  deal with the vector fields of the local coordinate basis $(H_A,H_p,L_{\alpha})$.
It can be shown that the time derivative of the function $\varphi$ which is  given on the path $(Q^{\ast}{}^A(t),\tilde f^p(t),a^{\alpha}(t))$ is  calculated in accordance with (\ref{f_dt}):

\begin{equation}
 \frac{d \varphi(Q^{\ast}{}^A(t),\tilde f^p(t),a^{\alpha}(t))}{dt}= 
 {\omega}^E H_E(\varphi)+{\omega}^pH_p({\varphi})+{\omega}^{\alpha} L_{\alpha}(\varphi),
\label{dif_gen_dt}
\end{equation}
where quasi-velocities ${\omega}^E$, ${\omega}^p$ and ${\omega}^{\alpha}$ are defined by (\ref{veloc_omega}).  By $H_E(\varphi)$ and $H_p(\varphi)$, we denote the action of the vector fields $H_E$ and $H_p$ on the function $\varphi$. A similar notation is used for $L_{\alpha}(\varphi)$.

First we consider the differential relation  between  $\omega ^A$ and $w^A$. Using  (\ref{dif_gen_dt}) for $Q^{\ast}{}^A(t)$, we get
\[
\frac{d {Q^{\ast}}^A(t)}{dt}=  
{\omega}^E H_E^A(Q^{\ast}(t)),
\]
where we write $H_E^A(Q^{\ast})$ for $H_E({Q^{\ast}}^A)$ which is equal to $N^A_E(Q^{\ast})$.

The obtained  equality for the time derivative of $Q^{\ast}(t)$ can be extended  to the corresponding equality for 
the deformed paths ${Q^{\ast}}^A(u,t)$:
\begin{equation}
  \frac{\partial {Q^{\ast}}^A(u,t)}{\partial t}=  H_E({Q^{\ast}}^A(u,t))\,{\omega}^E(u,t),
\label{difQ_t}
\end{equation}
where  ${\omega}^E(u,t)$ is given by the linear function of the velocities that are defined    for 
$Q^{\ast}{}^E(u,t)$.

On the other hand, for the partial derivative of $Q^{\ast}{}^A(u,t)$ with respect to $u$, 
we have, as supposed by the Poincar\'{e} method,
the following equation:
\begin{equation}
  \frac{\partial {Q^{\ast}}^A(u,t)}{\partial u}=
  H_E({Q^{\ast}}^A(u,t))\,{w}^E(u,t).
\label{difQ_u}
\end{equation}
Note that  introduced  variations ${w}^E(u,t)$ are independent inside of the time interval $(t_1,t_2)$ and vanish on its boundary, i.e. ${w}^E(u,t_1)={w}^E(u,t_2)=0$ . As functions, these variations   depend (functionally) on  deformations of the paths $Q^{\ast}{}^A(u,t)$. 

Now taking the partial derivative of (\ref{difQ_t}) with respect to $u$, we obtain  
\begin{eqnarray}
 \frac{\partial}{\partial u}\frac{\partial {Q^{\ast}}^A(u,t)}{\partial t}&=&\frac{\partial H^A_E(Q^{\ast})}{\partial {Q^{\ast}}^B}\,\frac{\partial {Q^{\ast}}^B}{\partial u}\, {\omega}^E+H^A_E\,\frac{\partial {\omega}^E(u,t)}{\partial u}
\nonumber\\
&=&\frac{\partial H^A_E}{\partial {Q^{\ast}}^B}\,H^B_Pw^P{\omega}^E+H^A_E\frac{\partial {\omega}^E}{\partial u}.
\label{difQ_t_u}
\end{eqnarray}
While, differentiation of  (\ref{difQ_u}) with respect to $t$ yields
\begin{equation}
 \frac{\partial}{\partial t}\frac{\partial {Q^{\ast}}^A(u,t)}{\partial u}=\frac{\partial H^A_E(Q^{\ast})}{\partial {Q^{\ast}}^B}\,
H^B_P{\omega}^Pw^E+H^A_E\frac{\partial {w}^E}{\partial t}.
\label{difQ_u_t}
\end{equation}
Subtracting (\ref{difQ_u_t}) from (\ref{difQ_t_u}), we obtain
\[
\Bigl( \frac{\partial H^A_E}{\partial {Q^{\ast}}^B}H^B_P-\frac{\partial H^A_P}{\partial {Q^{\ast}}^B}H^B_E\Bigr)\,{\omega}^Ew^P+H^A_{E'}\frac{\partial {\omega}^{E'}}{\partial u}-H^A_{E'}\frac{\partial {w}^{E'}}{\partial t}=0.
\]
 Taking 
into account  the commutation relation  
(\ref{commrelat_AB}) and the used   notation, by which $H^A_R(Q^{\ast})=N^A_R(Q^{\ast})$, we come to the following differential relation: 
\begin{equation}
 N^A_R\Bigl(\frac{\partial {\omega}^{R}}{\partial u}-\frac{\partial {w}^{R}}{\partial t}+{\mathbb C}^{R}_{PE}\,{\omega}^{E} {w}^{P}\Bigr)=0,
\label{relat_A_w}
\end{equation}
with ${\mathbb C}^R_{P E}=({\Lambda}^{\gamma}_PN^S_E-{\Lambda}^{\gamma}_EN^S_P)K^R_{\gamma,S}$.

Next we  derive the differential relation between the quasi-velocity $\omega ^p$ and  the variation $w^p$. It is done as in the previous case. 

For the path $\tilde f^p(t)$,  equation (\ref{dif_gen_dt}) is written as follows:
 \[
 \frac{d \tilde f^p(t)}{dt}=H_E(\tilde f^p(t)){\omega}^E+H_m(\tilde f^p(t)){\omega}^m=N^p_E(\tilde f^p(t))\,{\omega}^E+{\omega}^p.
\]
Provided that the deformations of the paths $Q^{\ast}{}^A(u,t)$ and $\tilde f(u,t)$ are properly chosen, we may obtain an analogouse representation for time derivative of  $\tilde f(u,t)$:  
 \begin{equation}
 \frac{\partial \tilde f^p(u,t)}{\partial t}=
N^p_E(\tilde f(u,t))\,{\omega}^E(u,t)+{\omega}^p(u,t),
\label{dif_f_dt}
\end{equation}
where now each ${\omega}^p(u,t)$ is a linear functions of the velocities defined  for  the deformed paths $\tilde f^m(u,t)$.

For the partial derivative of $\tilde f(u,t)$ with respect to $u$, we   take the  representation which is similar in form with (\ref{dif_f_dt}), but where the quasi-velocities of the deformed paths are replaced  by the corresponding deformations:
\begin{eqnarray}
 \frac{\partial \tilde f^p(u,t)}{\partial u}&=&H_R(\tilde f^p(u,t))w^R(u,t)+H_q(\tilde f^p(u,t))w^q(u,t)\nonumber\\
&=&N^p_R(\tilde f(u,t))\,w^R(u,t)+w^p(u,t).
\label{dif_f_du}
\end{eqnarray}
Taking the partial derivative of (\ref{dif_f_dt}) with respect to $u$, and then 
the partial derivative of (\ref{dif_f_du}) with respect to $t$, we get two equal expression:
\[
 \frac{{\partial}^2 \tilde f^p}{\partial u\, \partial t}=H_R (N^p_E)\,w^R {\omega}^E+H_n(N^p_E)w^n{\omega}^E+N^p_E
\,\frac{\partial {\omega}^E}{\partial u}+\frac{\partial {\omega}^p}{\partial u}
\]
and
\[
 \frac{{\partial}^2 \tilde f^p}{\partial t\, \partial u}=H_E (N^p_R)\,w^R {\omega}^E+H_n(N^p_R)\, {\omega}^nw^R+N^p_E
\,\frac{\partial w^E}{\partial t}+\frac{\partial w^p}{\partial t}.
\]
Subtracting the second expression  from the first, we obtain
\begin{eqnarray*}
&&(H_R (N^p_E)-H_E (N^p_R))w^R{\omega}^E+H_q(N^p_E)w^q{\omega}^E-H_q(N^p_R){\omega}^qw^R 
\nonumber\\
&&\;\;\;\;+N^p_E\Bigl(\frac{\partial {\omega}^E}{\partial u}
-\frac{\partial w^E}{\partial t}\Bigr)+\Bigl(\frac{\partial {\omega}^p}{\partial u}-\frac{\partial w^p}{\partial t}\Bigr)=0.
\nonumber\\
\end{eqnarray*}
Now because of $H_E(\tilde f^p)=N^p_E(\tilde f)$, where  $N^p_E(\tilde f)=-K^p_{\alpha}(\tilde f){\Lambda}^{\alpha}_E$ and   $K^p_{\alpha}(\tilde f)=({\bar J}_{\alpha})^p_m\tilde f^m$, the obtained equation can  be rewritten as follows:
\begin{eqnarray*}
&&[H_R,H_E](f^p)w^R{\omega}^E+H_n(N^p_E)w^n{\omega}^E-H_n(N^p_R){\omega}^nw^R 
\nonumber\\
&&\;\;\;\;+N^p_E\Bigl(\frac{\partial {\omega}^E}{\partial u}-\frac{\partial w^E}{\partial t}\Bigr)+\Bigl(\frac{\partial {\omega}^p}{\partial u}-\frac{\partial w^p}{\partial t}\Bigr)=0
\nonumber\\
\end{eqnarray*}
Then, replacing the commutator by its explicit expression from (\ref{commrelat_AB}),
we get the following differential relation:
\begin{eqnarray}
&&N^p_T\Bigl( \frac{\partial {\omega}^T}{\partial u}-\frac{\partial w^T}{\partial t}-{\mathbb C}^T_{ER}{\omega}^Ew^R\Bigr)
\nonumber\\
&&+\Bigl(\frac{\partial {\omega}^p}{\partial u}-\frac{\partial w^p}{\partial t} -{\mathbb C}^p_{ER}\,{\omega}^Ew^R   
-{\mathbb C}^p_{Eq}\,{\omega}^Ew^q-{\mathbb C}^p_{qR}\,{\omega}^qw^R
\Bigr)=0.
\label{relat_p_w}
\end{eqnarray}
We note that this differential relation can be written by using the condenced notation for indices:
\[
{\tilde N}^{\tilde A}_{\tilde B}\, \frac{\partial {\omega}^{\tilde B}}{\partial u}={\tilde N}^{\tilde A}_{\tilde B} \,\frac{\partial {w}^{\tilde B}}{\partial t}+{\tilde N}^{\tilde A}_{\tilde B}\,{\mathbb C}^{\tilde B}_{\tilde E \tilde R}\,{\omega}^{\tilde E}w^{\tilde R}.
\]

The last differential relation dealing with  the  partial derivatives of $\omega^{\alpha}$ and  
$w^\alpha$ is obtained  in  Appendix A. The result  
is given by 
equation (\ref{relat_a_w}):
\begin{eqnarray*}
&&\frac{\partial{\omega}^{\beta}}{\partial u}-\frac{\partial{w}^{\beta}}{\partial t}+{\mathbb C}^{\beta}_{ R  E}w^R{\omega}^{E}
 -   {\mathbb C}^{\beta}_{ E p}w^p{\omega}^{E}
 +{\mathbb C}^{\beta}_{R m}  w^R{\omega}^{m}
\nonumber\\
&&\;\;\;\;\;\;\;\;\;\;\;\;\;\;\;\;\;\;+{\mathbb C}^{\beta}_{p m}w^p{\omega}^m + {c}^{\beta}_{\nu \mu} w^{\nu} {\omega}^{\mu}=0.\nonumber\\
\end{eqnarray*}
Now we can proceed to derivation of the Lagrange-Poincar\'{e} equations.

\section{The Lagrange-Poincar\'{e} equations}
We make use of the Poincar\'{e} variational principle to derive the Lagrange-Poincar\'{e} equations. The variational integral, which we take for this purpose, is given by the functional (\ref{action}).
The variation $\delta S$ of this functional is defined by (\ref{var_funk}). It follows that to obtain  $\delta S$, first  we must   compute the derivative of the functional $S$ with respect to that  variable of the deformed paths which is related to the distortion of  the paths:

\begin{eqnarray}
&&\frac{d S}{d u}=\int_{t_1}^{t_2}\Bigl(\frac{\partial {\hat{\mathcal L}}}{\partial {\omega}^C}\frac{\partial {\omega}^C}{\partial u}+\frac{\partial {\hat{\mathcal L}}}{\partial {\omega}^p}\frac{\partial {\omega}^p}{\partial u}+\frac{\partial {\hat{\mathcal L}}}{\partial {\omega}^{\alpha}}\frac{\partial {\omega}^{\alpha}}{\partial u}+\frac{\partial {\hat{\mathcal L}}}{\partial Q^{\ast}{}^B}\frac{\partial Q^{\ast}{}^B}{\partial u}
\nonumber\\
   &&\;\;\;\;\;\;\;\;\;\;\;\;\;\;\;\;   +\frac{\partial {\hat{\mathcal L}}}{\partial {\tilde f}^{p}}\frac{\partial {\tilde f}^{p}}{\partial u}
+\frac{\partial {\hat{\mathcal L}}}{\partial {a}^{\alpha}}\frac{\partial {a}^{\alpha}}{\partial u}\Bigr)dt.
\label{dS_du}
\end{eqnarray}
In order to perform the integration by parts in the integral (\ref{dS_du}), we have to transform the terms with the  derivatives of the quasi-velocities of the integrand into new terms with the time derivatives of the variations. This can be made with the help of the obtained differential relations between the quasi-velocities and deformations. We begin with the transformation of the first two terms of the integrand.

Since in our case ${\tilde G}^{\rm H}_{Ap}={\tilde G}^{\rm H}_{pA}$, these terms can be rewritten as follows:
\begin{eqnarray}
&&
\frac{\partial {\hat{\mathcal L}}}{\partial {\omega}^C}\frac{\partial {\omega}^C}{\partial u}+\frac{\partial {\hat{\mathcal L}}}{\partial {\omega}^p}\frac{\partial {\omega}^p}{\partial u}= 
\nonumber\\
&&{\tilde G}^{\rm H}_{CD}\,\frac{\partial {\omega}^C}{\partial u}{\omega}^D+
{\tilde G}^{\rm H}_{Ap}\,\frac{\partial {\omega}^A}{\partial u}{\omega}^p
+{\tilde G}^{\rm H}_{Ap}\,{\omega}^A\frac{\partial {\omega}^p}{\partial u}
+{\tilde G}^{\rm H}_{pq}\,{\omega}^p\frac{\partial {\omega}^q}{\partial u}.
\label{X}
\end{eqnarray}

Denoting temporary the differential relation (\ref{relat_A_w}) by $(A)$, and (\ref{relat_p_w}) by $(B)$, let us consider their linear combination $ {\tilde G}^{\rm H}_{BA} \cdot (A)+ {\tilde G}^{\rm H}_{Bm}\cdot (B)=0$. Using the identity
\[
 {\tilde G}^{\rm H}_{BA}N^A_T+{\tilde G}^{\rm H}_{Bm}N^m_T={\tilde G}^{\rm H}_{BT}\;\;\;\;({\rm or}\;\;\; {\tilde G}^{\rm H}_{B\tilde A}N^{\tilde A}_T={\tilde G}^{\rm H}_{BT}),
\]
we  get the following differential relation: 
\begin{itemize}
\item[(A')]
\[
{\tilde G}^{\rm H}_{BT}\,\frac{\partial {\omega}^T}{\partial u}={\tilde G}^{\rm H}_{BT}\,\Bigl(\frac{\partial {w}^T}{\partial t}+{\mathbb C}^T_{ER}\,{\omega}^Ew^R\Bigr)
\]
\[-{\tilde G}^{\rm H}_{Bm}\Bigl(\frac{\partial {\omega^m}}{\partial u}-\frac{\partial {w^m}}{\partial t}-{\mathbb C}^m_{EF}{\omega}^Ew^F -   {\mathbb C}^m_{Eq} {\omega}^Ew^q-   {\mathbb C}^m_{qR}{\omega}^qw^R\Bigr).
\] 
\end{itemize}
Similarly, considering the linear combination $ {\tilde G}^{\rm H}_{pA} \cdot (A)+ {\tilde G}^{\rm H}_{pq}\cdot (B)=0$, and taking into account  
\[
{\tilde G}^{\rm H}_{pA}N^A_T+{\tilde G}^{\rm H}_{pq}N^q_T={\tilde G}^{\rm H}_{pT}\;\;\;\;({\rm or}\;\;\; {\tilde G}^{\rm H}_{p\tilde B  }N^{\tilde B}_T={\tilde G}^{\rm H}_{pT}),
\]
we obtain
\begin{itemize}
\item[(B')]
\[
 {\tilde G}^{\rm H}_{pT}\,\frac{\partial {\omega}^T}{\partial u}={\tilde G}^{\rm H}_{pT}\,\Bigl(\frac{\partial {w}^T}{\partial t}+{\mathbb C}^T_{ER}\,{\omega}^Ew^R\Bigr)
\]
\[-{\tilde G}^{\rm H}_{pm}\Bigl(\frac{\partial {\omega^m}}{\partial u}-\frac{\partial {w^m}}{\partial t}-{\mathbb C}^m_{EF}{\omega}^Ew^F -   {\mathbb C}^m_{Eq} {\omega}^Ew^q-   {\mathbb C}^m_{qR}{\omega}^qw^R\Bigr).
\]
\end{itemize}

 Multiplying (A') by ${\omega}^B$ and using the result of the multiplication  in the right hand side of (\ref{X}),
 we get for it the following expression:
\begin{eqnarray*} 
&&{\tilde G}^{\rm H}_{BT}\,{\omega}^B\Bigl(\frac{\partial {w}^T}{\partial t}+{\mathbb C}^T_{ER}\,{\omega}^Ew^R\Bigr)
\nonumber\\
&&-{\tilde G}^{\rm H}_{Bm}\,{\omega}^B\Bigl(\underline{\frac{\partial {\omega^m}}{\partial u}}-\frac{\partial {w^m}}{\partial t}-{\mathbb C}^m_{EF}{\omega}^Ew^F -   {\mathbb C}^m_{Eq} {\omega}^Ew^q-   {\mathbb C}^m_{qR}{\omega}^qw^R\Bigr)
\nonumber\\
&&+{\tilde G}^{\rm H}_{Ap}\,\frac{\partial {\omega}^A}{\partial u}{\omega}^p
+\underline{{\tilde G}^{\rm H}_{Ap}\,{\omega}^A\frac{\partial {\omega}^p}{\partial u}}
+{\tilde G}^{\rm H}_{pq}\,{\omega}^p\frac{\partial {\omega}^q}{\partial u}.
\end{eqnarray*}
We see that underlined terms are cancelled. Next we  insert  the result of the multiplication  (B') by ${\omega}^p$ in just obtained expression. As a consequence, we come to the following expression for the right hand side of (\ref{X}):
\begin{eqnarray}
 &&{\tilde G}^{\rm H}_{BT}\,{\omega}^B\Bigl(\frac{\partial {w}^T}{\partial t}+{\mathbb C}^T_{ER}\,{\omega}^Ew^R\Bigr)
\nonumber\\
&&-{\tilde G}^{\rm H}_{Bm}\,{\omega}^B\Bigl(-\frac{\partial {w^m}}{\partial t}-{\mathbb C}^m_{EF}{\omega}^Ew^F -   {\mathbb C}^m_{Eq} {\omega}^Ew^q-   {\mathbb C}^m_{qR}{\omega}^qw^R\Bigr)
\nonumber\\
&&+ {\tilde G}^{\rm H}_{pT}\,{\omega}^p\Bigl(\frac{\partial {w}^T}{\partial t}+{\mathbb C}^T_{ER}\,{\omega}^Ew^R\Bigr)-{\tilde G}^{\rm H}_{pm}\,{\omega}^p\Bigl(\underline{\frac{\partial {\omega^m}}{\partial u}}-\frac{\partial {w^m}}{\partial t}
\nonumber\\
&&-{\mathbb C}^m_{EF}{\omega}^Ew^F -   {\mathbb C}^m_{Eq} {\omega}^Ew^q-   {\mathbb C}^m_{qR}{\omega}^qw^R\Bigr)
+\underline{{\tilde G}^{\rm H}_{pq}\,{\omega}^p\frac{\partial {\omega}^q}{\partial u}}.
\label{X+1}
\end{eqnarray}
Since  the underlined terms in this expression are  cancelled, now we can perform  the integration by parts of this expression.

Before writing out the result of the integration, it worth to note  that 
one can rewrite the obtained expression in a compact form
by making use of  the condensed notation  in which $Q^{\ast}{}^{\tilde A}$ means $(Q^{\ast}{}^{A},\tilde f^p)$: 
\begin{eqnarray*}
&&{\tilde G}^{\rm H}_{{\tilde B}T}\,{\omega}^{\tilde B}\Bigl(\frac{\partial {w}^T}{\partial t}+{\mathbb C}^T_{ER}\,{\omega}^Ew^R\Bigr) 
\nonumber\\
&&+{\tilde G}^{\rm H}_{{\tilde B}m}\,{\omega}^{\tilde B}\Bigl(\frac{\partial {w^m}}{\partial t}+{\mathbb C}^m_{EF}{\omega}^Ew^F +   {\mathbb C}^m_{Eq} {\omega}^Ew^q+   {\mathbb C}^m_{qR}{\omega}^qw^R\Bigr)
\nonumber\\
&&={\tilde G}^{\rm H}_{{\tilde B}{\tilde T}}\,{\omega}^{\tilde B}\Bigl(\frac{\partial {w}^{\tilde T}}{\partial t}+{\mathbb C}^{\tilde T}_{{\tilde E}{\tilde R}}\,{\omega}^{\tilde E}w^{\tilde R}\Bigr). 
\end{eqnarray*}
 In our case ${\mathbb C}^T_{Eq}=0,\;\;{\mathbb C}^T_{qR}=0,\;\;{\mathbb C}^T_{qp}=0$  and ${\mathbb C}^m_{qp}=0.$)

The result of the integration by parts of (\ref{X+1}) is given as follows:
\begin{eqnarray*}
 &&\int^{t_2}_{t_1}\Bigl(\frac{\partial {\hat{\mathcal L}}}{\partial {\omega}^C}\frac{\partial {\omega}^C}{\partial u}
+\frac{\partial {\hat{\mathcal L}}}{\partial {\omega}^p}\frac{\partial {\omega}^p}{\partial u}\Bigr)dt
=
\Bigl({\tilde G}^{\rm H}_{BT}\,{\omega}^B{w}^{T}+{\tilde G}^{\rm H}_{pT}\,{\omega}^p{w}^{T}\Bigr)\Bigl|^{t_2}_{t_1}
\nonumber\\
&&-\int^{t_2}_{t_1}\Bigl(\frac{d}{dt}\Bigl({\tilde G}^{\rm H}_{BT}\,{\omega}^B+{\tilde G}^{\rm H}_{pT}\,{\omega}^p\Bigr){w}^{T}- 
{\tilde G}^{\rm H}_{BT}\,{\omega}^B\,{\mathbb C}^{T}_{ER}\,{\omega}^E w^R\nonumber\\
&&-{\tilde G}^{\rm H}_{pT}\,{\omega}^p\,{\mathbb C}^{T}_{ER}\,{\omega}^E w^R
-{\tilde G}^{\rm H}_{Bm}\,{\omega}^B\,({\mathbb C}^{m}_{EF}\,{\omega}^E w^F
+{\mathbb C}^{m}_{qR}\,{\omega}^q w^R)
\nonumber\\
&&-{\tilde G}^{\rm H}_{pm}\,{\omega}^p\,({\mathbb C}^{m}_{EF}\,{\omega}^E w^F
+{\mathbb C}^{m}_{qR}\,{\omega}^q w^R)
\Bigr)dt
\nonumber\\
&&+\Bigl({\tilde G}^{\rm H}_{Bm}\,{\omega}^B{w}^{m}+{\tilde G}^{\rm H}_{pm}\,{\omega}^p{w}^{m}\Bigr)\Bigl|^{t_2}_{t_1}
\nonumber\\
&&-\int^{t_2}_{t_1}\Bigl(\frac{d}{dt}\Bigl({\tilde G}^{\rm H}_{Bm}\,{\omega}^B+{\tilde G}^{\rm H}_{pm}\,{\omega}^p\Bigr){w}^{m}- 
{\tilde G}^{\rm H}_{Bm}\,{\omega}^B\,{\mathbb C}^{m}_{Eq}\,{\omega}^E w^q\nonumber\\
&&-{\tilde G}^{\rm H}_{pm}\,{\omega}^p\,{\mathbb C}^{m}_{Eq}\,{\omega}^E w^q
\Bigr)dt.
\end{eqnarray*}
The right hand side of the obtained expression can be rewritten as
\begin{eqnarray*}
 &&\!\!\!\!\!\!\!\Bigl({\tilde G}^{\rm H}_{BT}\,{\omega}^B{w}^{T}+{\tilde G}^{\rm H}_{pT}\,{\omega}^p{w}^{T}\Bigr)\Bigl|^{t_2}_{t_1}
\nonumber\\
&&\!\!\!\!\!\!\!\!\!\!\!\!\!\!-\int^{t_2}_{t_1}\!dt\Bigl(\frac{d}{dt}\Bigl(\frac{\partial {\hat{\mathcal L}}}{\partial {\omega}^{R}}\Bigr)- 
\Bigl(\frac{\partial {\hat{\mathcal L}}}{\partial {\omega}^{T}}\Bigr){\mathbb C}^{T}_{ER}
{\omega}^E 
-\Bigl(\frac{\partial {\hat{\mathcal L}}}{\partial {\omega}^{m}}\Bigr)({\mathbb C}^{m}_{ER}{\omega}^E 
+{\mathbb C}^{m}_{qR}{\omega}^q)\Bigr) w^R
\nonumber\\
&&\!\!\!\!\!\!\!\!\!\!\!\!\!\!+\Bigl({\tilde G}^{\rm H}_{Bm}\,{\omega}^B{w}^{m}+{\tilde G}^{\rm H}_{pm}\,{\omega}^p{w}^{m}\Bigr)\Bigl|^{t_2}_{t_1}
-\int^{t_2}_{t_1}\!\!dt\Bigl(\frac{d}{dt}\Bigl(\frac{\partial {\hat{\mathcal L}}}{\partial {\omega}^{q}}\Bigr)- 
\Bigl(\frac{\partial {\hat{\mathcal L}}}{\partial {\omega}^{q}}\Bigr){\mathbb C}^{m}_{Eq}\,{\omega}^E\Bigr) w^q.\nonumber\\
\end{eqnarray*}

In a similar manner we can carry out  the integration  of the following term in the integral (\ref{X}). This lead us to 
\begin{eqnarray*}
&&\int^{t_2}_{t_1}
\frac{\partial {\hat{\mathcal L}}}{\partial {\omega}^{\epsilon}}\frac{\partial {\omega}^{\epsilon}}{\partial u}dt=
\bigl({\tilde d}_{\epsilon \nu}\,{\omega}^{\nu}{w}^{\epsilon}\bigr)\Bigl|^{t_2}_{t_1}
\nonumber\\
&&-\int^{t_2}_{t_1}\Bigl(\frac{d}{dt}\Bigl(\frac{\partial{\hat{\mathcal L}}}{\partial {\omega}^{\epsilon}}\Bigr){w}^{\epsilon} 
-\Bigl(\frac{\partial{\hat{\mathcal L}}}{\partial {\omega}^{\epsilon}}\Bigr)
(-{\mathbb C}^{\epsilon}_{RE}\,{\omega}^E -{\mathbb C}^{\epsilon}_{Rm}\,{\omega}^m) w^R
\nonumber\\
&&-\Bigl(\frac{\partial{\hat{\mathcal L}}}{\partial {\omega}^{\epsilon}}\Bigr)({\mathbb C}^{\epsilon}_{Ep}\,{\omega}^E -{\mathbb C}^{\epsilon}_{pm}\,{\omega}^m) w^p+\Bigl(\frac{\partial{\hat{\mathcal L}}}{\partial {\omega}^{\epsilon}}\Bigr){ c}^{\epsilon}_{\nu \mu}\,{\omega}^{\mu} w^{\nu}
\Bigr)dt.
\end{eqnarray*}
The remaining terms of the integrand in the integral (\ref{dS_du}) can be transformed as
\begin{eqnarray*}
 &&\!\!\!\!\!\!\frac{\partial {\hat{\mathcal L}}}{\partial Q^{\ast}{}^B}\frac{\partial Q^{\ast}{}^B}{\partial u}
+\frac{\partial {\hat{\mathcal L}}}{\partial {\tilde f}^p}\frac{\partial {\tilde f}^{p}}{\partial u}
+\frac{\partial {\hat{\mathcal L}}}{\partial {a}^{\alpha}}\frac{\partial {a}^{\alpha}}{\partial u}=
\frac{\partial {\hat{\mathcal L}}}{\partial Q^{\ast}{}^B}N^B_Ew^E+\frac{\partial {\hat{\mathcal L}}}{\partial {\tilde f}^p}N^p_Ew^E+\frac{\partial {\hat{\mathcal L}}}{\partial {\tilde f}^p}w^p
\nonumber\\
&&+\frac{\partial {\hat{\mathcal L}}}{\partial {a}^{\alpha}}\bigl(-N^C_E\tilde\mathscr A^{\beta}_Cv^{\alpha}_{\beta}w^E-N^q_E\tilde\mathscr A^{\mu}_qv^{\alpha}_{\mu}w^E-\tilde\mathscr A^{\sigma}_pv^{\alpha}_{\sigma}w^p+v^{\alpha}_{\beta}w^{\beta}\bigr)=
\nonumber\\
&&H_E({\hat{\mathcal L}})w^E+H_p({\hat{\mathcal L}})w^p+L_{\alpha}({\hat{\mathcal L}})w^{\alpha}.
\end{eqnarray*}

Since the variations $w^E$, $w^p$ and $w^{\alpha}$ are independent between themselves and satisfy   the  standard boundary conditions, this enables  us to obtain the following system of the Lagrange-Poincar\'{e} equations:
\begin{eqnarray}
&&-\frac{d}{dt}\Bigl(\frac{\partial{\hat{\mathcal L}}}{\partial {\omega}^E}\Bigr)+\Bigl(\frac{\partial{\hat{\mathcal L}}}{\partial {\omega}^T}\Bigr){\mathbb C}^{T}_{CE}\,{\omega}^C+\Bigl(\frac{\partial{\hat{\mathcal L}}}{\partial {\omega}^p}\Bigr)({\mathbb C}^{p}_{CE}\,{\omega}^C+{\mathbb C}^{p}_{qE}\,{\omega}^q)
\nonumber\\
&&\;\;\;\;\;\;\;+\Bigl(\frac{\partial{\hat{\mathcal L}}}{\partial {\omega}^{\alpha}}\Bigr) ({\mathbb C}^{\alpha}_{CE}\,{\omega}^C+{\mathbb C}^{\alpha}_{mE}\,{\omega}^m)+H_E(\hat{\mathcal L})=0
\label{eq_Poinc_Q}
\end{eqnarray}

\begin{eqnarray}
 &&\!\!\!\!\!\!\!\!\!\!\!\!\!\!\!\!\!\!\!\!\!\!\!\!-\frac{d}{dt}\Bigl(\frac{\partial{\hat{\mathcal L}}}{\partial {\omega}^m}\Bigr)+\Bigl(\frac{\partial{\hat{\mathcal L}}}{\partial {\omega}^p}\Bigr){\mathbb C}^{p}_{Em}\,{\omega}^E
\nonumber\\
&&\!\!\!\!\!\!\!\!\!+\Bigl(\frac{\partial{\hat{\mathcal L}}}{\partial {\omega}^{\alpha}}\Bigr) ({\mathbb C}^{\alpha}_{Em}\,{\omega}^E+{\mathbb C}^{\alpha}_{pm}\,{\omega}^p)+H_m(\hat{\mathcal L})=0
\label{eq_Poinc_f}
\end{eqnarray}
\begin{equation}
\!\!\!\!\!\!\!\!\!\!\!\!\!\!\!\!\!\!\!\!\!\!\!\!\!\!\!\!\!\!\!\!\!\!\!\!\!-\frac{d}{dt}\Bigl(\frac{\partial{\hat{\mathcal L}}}{\partial {\omega}^{\alpha}}\Bigr)+\Bigl(\frac{\partial{\hat{\mathcal L}}}{\partial {\omega}^{\beta}}\Bigr){c}^{\beta}_{\mu \alpha}\,{\omega}^{\mu}+L_{\alpha}(\hat{\mathcal L})=0.
\label{eq_Poinc_alfa}
\end{equation}
The first two equations of this system are   the horizontal equations, and the last equation, describing the motion of the group variable, is the   vertical one.

\section{Concluding remarks}

In the paper, we have obtained the Lagrange-Poincar\'{e} equations using the dependent variables, determined on a global surface $\tilde \Sigma$. This means that the principal fiber bundle related to our original mechanical system is  trivial one.  One meets with a similar case in gauge theories, where it is not possible in general to obtain a set of  gauge-invariant variables that are globally determined on the orbit space of the principal fiber bundle.  

We note that obtained  equations may be also  used  for description of the local evolution (in terms of dependent variables) given  on a appropriate domain of the orbit space of the non-trivial 
principal bundle. But, as in gauge theories, the problem of ``gluing'' these evolutions into the global one is not yet settled.

We remark that our horizontal equations are analogous in form with that ones from \cite{Marsden_2}. 
But the ``structure constants'' in our case are calculated for the horizontal lift basis and differ from the structure constants of the cited  work. 

Note also that the horizontal  Lagrange-Poincar\'{e} equations of the present case can be derived from the similar equations of our paper \cite{Storchak_eq_Poinc} by considering them as if they were written in terms of  the  condensed notations, that have been used in this paper. However, the ``structure constants'' should be  taken those as in (\ref{commrelat_AB}), (\ref{commrelat_Ap}) and (\ref{commrelat_pq}).

\appendix
\section*{Appendix A}
\section*{Differential relation  between   $\omega ^{\alpha}$ and  $w^{\alpha}$}
\setcounter{equation}{0}
\def\theequation{A.\arabic{equation}}

By the general formula (\ref{dif_gen_dt})   applied to the path $ a^\alpha(t)$,
the velocity $d a^\alpha(t)/{dt}$  is decomposed as follows:
\[
 \frac{d a^{\alpha}(t)}{d t}=H_E(a^{\alpha}(t)){\omega}^E(t)+H_m(a^{\alpha}(t)){\omega}^m(t) +L_{\mu}(a^{\alpha}(t)){\omega}^{\mu}(t),
\]
where each of the  quasi-velocities ${\omega}^E(t)$, ${\omega}^m(t)$ and ${\omega}^{\mu}(t)$ is a linear function of the velocities.
Taking  suffient small variations of the paths, such a representation for $d a^\alpha(t)/{dt}$ can be extended to the representation which determines the decomposition of the velocity 
$d a^\alpha(u,t)/{dt}$: \footnote{$a^\alpha(u,t)$ is a deformation of the original  path $a^\alpha(t)$, 
i.e., for instance, as $a^\alpha(u,t)=a^\alpha(t)+uW^{\alpha}(t)$.}
\begin{equation}
 \frac{\partial a^{\alpha}(u,t)}{\partial  t}=H_E(a^{\alpha}(u,t)){\omega}^E(u,t)+H_m(a^{\alpha}(u,t)){\omega}^m(u,t) +L_{\mu}(a^{\alpha}(u,t)){\omega}^{\mu}(u,t).
\label{difa_t}
\end{equation}
In this representation, the quasi-velocities ${\omega}^E(u,t)$, ${\omega}^m(u,t)$ and ${\omega}^{\mu}(u,t)$ denote the  linear functions of the velocities of the deformed paths.

In the considered calculus of variation, it is supposed  that $\frac{\partial a^{\alpha}(u,t)}{\partial  u}$  is defined similarly to (\ref{difa_t}), where, however, quasi-velocities are replaced by variations: 
\begin{equation}
 \frac{\partial a^{\alpha}(u,t)}{\partial  u}=H_R(a^{\alpha}(u,t)){w}^R(u,t)+H_p(a^{\alpha}(u,t)){w}^p(u,t) +L_{\nu}(a^{\alpha}(u,t)){w}^{\nu}(u,t).
\label{difa_u}
\end{equation}
Taking the partial derivative of (\ref{difa_t}) with respect to $u$, we get
\begin{eqnarray*}
\displaystyle
&&\frac{{\partial}^2 a^{\alpha}}{\partial u \partial t}=
H_RH_E(a^{\alpha})w^R{\omega}^E+H_RH_m(a^{\alpha})w^R{\omega}^m+H_RL_{\mu}(a^{\alpha})w^R{\omega}^{\mu}
\nonumber\\
&&+H_pH_E(a^{\alpha})w^p{\omega}^E+H_pH_m(a^{\alpha})w^p{\omega}^m+H_pL_{\mu}(a^{\alpha})w^p{\omega}^{\mu}
+L_{\nu}H_E(a^{\alpha})w^{\nu}{\omega}^{E}
\nonumber\\
&&+L_{\nu}H_m(a^{\alpha})w^{\nu}{\omega}^{m}+L_{\nu}L_{\mu}(a^{\alpha})w^{\nu}{\omega}^{\mu}
+H_E(a^{\alpha})\frac{\partial{\omega}^{E}}{\partial u}+H_m(a^{\alpha})\frac{\partial{\omega}^{m}}{\partial u}
\nonumber\\
&&+L_{\mu}(a^{\alpha})\frac{\partial{\omega}^{\mu}}{\partial u}.
\end{eqnarray*}
Similarly, the differentiation of (\ref{difa_u}) with respect to $t$ yields
\begin{eqnarray*}
 \frac{{\partial}^2 a^{\alpha}}{\partial t \partial u}= H_EH_R(a^{\alpha}){\omega}^Ew^R+H_EH_p(a^{\alpha}){\omega}^Ew^p+H_EL_{\nu}(a^{\alpha}){\omega}^{E}w^{\nu}
\end{eqnarray*}
\[
+H_mH_R(a^{\alpha}){\omega}^mw^R+H_mH_p(a^{\alpha}){\omega}^mw^p+H_mL_{\nu}(a^{\alpha}){\omega}^{m}w^{\nu}
+L_{\mu}H_R(a^{\alpha}){\omega}^{\mu}w^R
\]
\[
+L_{\mu}H_p(a^{\alpha}){\omega}^{\mu}w^R+L_{\mu}L_{\nu}(a^{\alpha}){\omega}^{\mu}w^{\nu}
+H_R(a^{\alpha})\frac{\partial w^{R}}{\partial t}+H_p(a^{\alpha})\frac{\partial{w}^{p}}{\partial u}+L_{\nu}(a^{\alpha})\frac{\partial{w}^{\nu}}{\partial t}.
\]
Subtracting this expression from the expression given above, we obtain;
\begin{eqnarray*}
  &&[H_R,H_E](a^{\alpha})w^R{\omega}^E+[H_p,H_E](a^{\alpha})w^p{\omega}^E+[H_R,H_m](a^{\alpha})w^R{\omega}^m
\end{eqnarray*}
\[
+[H_p,H_m](a^{\alpha})w^p{\omega}^m+[H_p,L_{\mu}](a^{\alpha})w^p{\omega}^{\mu}+[H_R,L_{\mu}](a^{\alpha})w^R{\omega}^{\mu}
\]
\[
+[L_{\nu},H_E](a^{\alpha})w^{\nu}{\omega}^{E}+[L_{\nu},H_m](a^{\alpha})w^{\nu}{\omega}^{m}+[L_{\nu},L_{\mu}](a^{\alpha})w^{\nu}{\omega}^{\mu}
\]
\[
+H_E(a^{\alpha})\Bigl(\frac{\partial{\omega}^{E}}{\partial u}-\frac{\partial{w}^{E}}{\partial t}\Bigr)+H_m(a^{\alpha})\Bigl(\frac{\partial{\omega}^{m}}{\partial u}-\frac{\partial{w}^{m}}{\partial u}\Bigr)+L_{\mu}(a^{\alpha})\Bigl(\frac{\partial{\omega}^{\mu}}{\partial u}-\frac{\partial{w}^{\mu}}{\partial t}\Bigr)=0.
\]

By making use of  the commutation relations (\ref{commrelat_AB}), (\ref{commrelat_Ap}), (\ref{commrelat_pq}), together with the commutation relations for the left-invariant vector fields $L_{\alpha}$, we rewrite the obtained equation  as follows:
\[
({\mathbb C}^{T}_{R  E}H_T(a^{\alpha})+{\mathbb C}^{p}_{ R  E}H_p(a^{\alpha})
+{\mathbb C}^{\gamma}_{ R  E}L_{\gamma}(a^{\alpha}))w^R{\omega}^{E}
\]
\[
-({\mathbb C}^{m}_{ E p}H_m(a^{\alpha})+   {\mathbb C}^{\gamma}_{ E p}L_{\gamma}(a^{\alpha}))w^p{\omega}^{E}
+{\mathbb C}^{\gamma}_{p m}L_{\gamma}(a^{\alpha})w^p{\omega}^{m}
\]
\[
+({\mathbb C}^{q}_{R m} H_q(a^{\alpha})+{\mathbb C}^{\gamma}_{R m}  L_{\gamma}(a^{\alpha}))w^R{\omega}^{m}
+ {c}^{\gamma}_{\nu \mu}   L_{\gamma}(a^{\alpha}) w^{\nu} {\omega}^{\mu}  
\]
\[
+H_E(a^{\alpha})\Bigl(\frac{\partial{\omega}^{E}}{\partial u}-\frac{\partial{w}^{E}}{\partial t}\Bigr)+H_m(a^{\alpha})\Bigl(\frac{\partial{\omega}^{m}}{\partial u}-\frac{\partial{w}^{m}}{\partial u}\Bigr)+L_{\mu}(a^{\alpha})\Bigl(\frac{\partial{\omega}^{\mu}}{\partial u}-)\frac{\partial{w}^{\mu}}{\partial t}\Bigr)=0.
\]

We note that
\[
 H_T(a^{\alpha})=-N^D_T{\tilde {\mathscr A}}^{\mu}_DL_{\mu}(a^{\alpha})-N^m_T{\tilde {\mathscr A}}^{\mu}_mL_{\mu}(a^{\alpha})=-(N^D_T{\tilde {\mathscr A}}^{\mu}_D+N^m_T{\tilde {\mathscr A}}^{\mu}_m)v^{\alpha}_{\mu}(a)
\]
and
$
 H_p(a^{\alpha})=-{\tilde {\mathscr A}}^{\mu}_pv^{\alpha}_{\mu}(a)$.
 Therefore, multiplying the equation by $u^{\beta}_{\alpha}(a)$, we 
 get rid off the common multiplier $v^{\alpha}_{\mu}(a)$.
 As a result, we arrive at
\[
 \bigl[-{\mathbb C}^{T}_{R  E}(N^D_T{\tilde {\mathscr A}}^{\beta}_D+N^m_T{\tilde {\mathscr A}}^{\beta}_m)-{\mathbb C}^{p}_{ R  E}{\tilde {\mathscr A}}^{\beta}_p
+{\mathbb C}^{\beta}_{ R  E}\bigr]w^R{\omega}^{E}
\]
\[
 -(-{\mathbb C}^{m}_{ E p}{\tilde {\mathscr A}}^{\beta}_m+   {\mathbb C}^{\beta}_{ E p})w^p{\omega}^{E}
+(-{\mathbb C}^{q}_{R m}{\tilde {\mathscr A}}^{\beta}_q  +{\mathbb C}^{\beta}_{R m}  )w^R{\omega}^{m}
\]
\[
+ {c}^{\beta}_{\nu \mu} w^{\nu} {\omega}^{\mu}+{\mathbb C}^{\beta}_{p m}w^p{\omega}^m -\bigl( N^D_T{\tilde {\mathscr A}}^{\beta}_D+N^m_T{\tilde {\mathscr A}}^{\beta}_m\bigr)
\Bigl(\frac{\partial{\omega}^{E}}{\partial u}-\frac{\partial{w}^{E}}{\partial t}\Bigr)
\]
\[
 -{\tilde {\mathscr A}}^{\beta}_m\Bigl(\frac{\partial{\omega}^{m}}{\partial u}-\frac{\partial{w}^{m}}{\partial u}\Bigr)
+\Bigl(\frac{\partial{\omega}^{\beta}}{\partial u}-\frac{\partial{w}^{\beta}}{\partial t}\Bigr)=0
\].

First we observe  that in just obtained equation, the sum of the terms with common multiplier  $N^D_T$  has to  vanish because of  the differential relation (\ref{relat_A_w}) 
for ${\omega}^{A}$. Hence,   we deal, in fact,  with the  equation which looks as follows:
$ {\tilde {\mathscr A}}^{\beta}_m(...) +(...)=0$.
But in this equation  the first summand  also  has to vanish because of the second differential identity (\ref{relat_p_w})  for ${\omega}^{m}$.
It follows that final equation representing the differential relation for ${\omega}^{\beta}$ is given by 
\begin{eqnarray}
 &&\frac{\partial{\omega}^{\beta}}{\partial u}-\frac{\partial{w}^{\beta}}{\partial t}+{\mathbb C}^{\beta}_{ R  E}w^R{\omega}^{E}
 -   {\mathbb C}^{\beta}_{ E p}w^p{\omega}^{E}
 +{\mathbb C}^{\beta}_{R m}  w^R{\omega}^{m}
\nonumber\\
&&\;\;\;\;\;\;\;\;\;\;\;\;\;\;\;\;\;+{\mathbb C}^{\beta}_{p m}w^p{\omega}^m + {c}^{\beta}_{\nu \mu} w^{\nu} {\omega}^{\mu}=0.
\label{relat_a_w}
\end{eqnarray}

\appendix
\section*{Appendix B}
\section*{Projectors, their properties and\\ some identities}
\setcounter{equation}{0}
\def\theequation{A.\arabic{equation}}

\subsection*{The horizontal projector $\Pi^{\tilde A}_{\tilde B}$}

By definition
$${\Pi}^{\tilde A}_{\tilde B}={\delta}^{\tilde A}_{\tilde B}-K^{\tilde A}_{\alpha}d^{\alpha \beta}K^{\tilde D}_{\beta}G_{\tilde D \tilde B}.$$
Its components are
$$\Pi^{\tilde A}_{\tilde B}=(\Pi^A_B,\Pi^A_n,\Pi^m_B,\Pi^m_n),$$

$${\Pi}^{A}_{B}={\delta}^A_B-K^{A}_{\alpha}d^{\alpha \beta}K^{D}_{\beta}G_{D B},\;\;\;\;\;
{\Pi}^{A}_{n}=-K^{A}_{\mu}d^{\mu \nu}K^{p}_{\nu}G_{pn},$$

$${\Pi}^{m}_{B}=-K^{m}_{\mu}d^{\mu \nu}K^{D}_{\nu}G_{D B},\;\;\;\;
{\Pi}^{m}_{n}={\delta}^m_n-K^{m}_{\mu}d^{\mu \nu}K^{r}_{\nu}G_{rn}.$$

The main properties:
\[
 \Pi^{\tilde A}_{\tilde B}\Pi^{\tilde B}_{\tilde C}=\Pi^{\tilde A}_{\tilde C}, \;\;\;
{\Pi}^{\tilde L}_{\tilde B}N^{\tilde A}_{\tilde L}=N^{\tilde A}_{\tilde B},\;\;\;
{\Pi}_{\tilde L}^{\tilde A}N_{\tilde C}^{\tilde L}={\Pi}_{\tilde C}^{\tilde A},\;\;\;
{\Pi}^{\tilde A}_{\tilde E}K^{\tilde E}_{\alpha}=0.
\]

Under the transformation $$Q^A=F^A(Q^{\ast},a), \;\; F^A_C(Q^{\ast},a)\equiv\frac{\partial F^A(Q,a)}{\partial Q^C}|_{Q=Q^{\ast}},\;\;  \check F^C_A\equiv F^C_A(F(Q{}^{\ast},a), a^{-1})$$
and because of 
\[
 K^B_{\alpha}(F(Q^{\ast},a))={\rho}^{\mu}_{\alpha}(a)K^D_{\mu}(Q^{\ast})F^B_D(Q^{\ast},a),\;
 K^p_{\alpha}(\bar D(a)\tilde f)={\rho}^{\mu}_{\alpha}(a)K^q_{\mu}(\tilde f)\bar D^p_q(a),
\]
\[
 G_{AB}(F(Q^{\ast},a))=G_{DC}(Q^{\ast})\check F^D_A\check F^C_B,\;\;
 G_{pq}=G_{mn}\bar D^m_p(a)\bar D^n_q(a),
\]
\[
 d_{\alpha\beta}(Q,f)={\rho}^{\mu}_{\alpha}(a){\rho}^{\nu}_{\beta}(a)\,d_{\mu\nu}(Q^{\ast},\tilde f),
\]
it follows that
\[
\Pi^A_B(Q,f)=F^A_C\,\Pi^C_D(Q^{\ast},\tilde f)\,\check F^D_B,\;\;\Pi^A_n(Q,f)= F^A_B\,\Pi^B_r(Q^{\ast},\tilde f)\, D^r_n(a),
\]
\[ 
\Pi^m_B(Q,f)=\check F^C_B\,\Pi^q_C(Q^{\ast},\tilde f)\,\bar D^m_q(a),\;\;
\Pi^m_n(Q,f)=\bar D^m_q(a)\,\Pi^q_s(Q^{\ast},\tilde f) D^s_n(a).
\]

\subsection*{The projector $N^{\tilde A}_{\tilde B}$}

Its components:
$$N^{\tilde A}_{\tilde C}=(N^A_B,N^A_n,N^m_B,N^m_n),$$ 

$$N^A_B={\delta}^A_B-K^A_{\mu}({\Phi}^{-1})^{\mu}_{\nu}\,{\chi}^{\nu}_B,\;\;
N^A_n=0,\;\;  N^m_B=-K^m_{\alpha}({\Phi}^{-1})^{\alpha}_{\mu}\,{\chi}^{\mu}_B,
\;\;  N^m_n={\delta}^m_n.$$
The main properties:
$$N^{\tilde A}_{\tilde B} N^{\tilde B}_{\tilde C}=N^{\tilde A}_{\tilde C},\;\;\;
(P_{\bot})^{\tilde L}_{\tilde B}\,N^{\tilde C}_{\tilde L}=(P_{\bot})^{\tilde C}_{\tilde B},\;\;\;      
N^{\tilde A}_{\tilde B}\,(P_{\bot})^{\tilde C}_{\tilde A}=N^{\tilde C}_{\tilde B}.
$$
Transformations
\[
 N^A_C(Q^{\ast})=F^B_C(Q^{\ast},a)N^M_B(F(Q^{\ast},a))\check F_M^A(Q^{\ast},a),\,N^A_C(Q^{\ast})\equiv N^A_C(F(Q^{\ast},e)),
\]
$e$ is the unity element of the group.

\subsection*{The projector $(P_{\bot})^{\tilde A}_{\tilde B}$}

Its components:
\[
 (P_{\bot})^{\tilde A}_{\tilde B}=(\,(P_{\bot})^A_B,\, (P_{\bot})^A_n,\, (P_{\bot})^m_B,\,(P_{\bot})^m_n\,),
\]
\[
 (P_\bot)^A_B=\delta^A_B-\chi ^{\alpha}_{B}\,(\chi \chi ^{\top})^{-1}{}^{\beta}_{\alpha}\,(\chi ^{\top})^A_{\beta},\;(\chi ^{\top})^A_{\mu}=G^{AB}{\gamma}_{\mu\nu}\chi ^{\nu}_B,\;{\gamma}_{\mu\nu}=K^A_{\mu}G_{AB}K^B_{\nu}, 
\]
 $(P_{\bot})^{A}_{n}=0,\;\;(P_{\bot})^{m}_{B}=0,\;\;(P_{\bot})^{m}_{n}=\delta ^m_n.$

\subsection*{Some identities derived from $K^{\tilde R}_{\gamma}{\tilde G}^H_{{\tilde R\tilde A} }=0$}

(1) $\tilde A\to A$
\[
 K^R_{\gamma}{\tilde G}^H_{RA}+K^p_{\gamma}{\tilde G}^H_{pA}=0\;\;\;\;\;{\rm or}\;\;\;K^{\tilde R}_{\gamma}{\tilde G}^H_{{\tilde R}A}=0
\]
\[
(A)\;\;\;\;K^R_{\gamma, D}{\tilde G}^H_{RA}+K^R_{\gamma}{\tilde G}^H_{RA,D}+K^p_{\gamma}{\tilde G}^H_{pA,D}=0.
\]
\[
\!\!\!\!(D)\;\;\;\;{\tilde G}^H_{AR,q}K^R_{\gamma}+{\tilde G}^H_{Ap,q}K^p_{\gamma}+{\tilde G}^H_{Ap}K^p_{\gamma,q}=0. 
\]

$\!\!\!\!\!\!\!(2)$ $\tilde A\to p$
\[
 {\tilde G}^H_{pq}K^q_{\mu}+{\tilde G}^H_{pA}K^A_{\mu}=0\;\;\;\;\;{\rm or}\;\;\;{\tilde G}^H_{p{\tilde R}}K^{\tilde R}_{\mu}=0
\]
\[
\!\!\!\!(B)\;\;\;\;{\tilde G}^H_{pR,n}K^R_{\mu}+{\tilde G}^H_{pr,n}K^r_{\mu}+{\tilde G}^H_{pr}K^r_{\mu,n}=0.
\]
\[
\!\!\!(C)\;\;\;\;{\tilde G}^H_{pR,D}K^R_{\mu}+{\tilde G}^H_{pr,D}K^r_{\mu}+{\tilde G}^H_{pR}K^R_{\mu,n}=0. 
\]
These relations are obtained as a result of the differentiations.

\subsection*{Killing relations for the horizontal metric $G^H_{\tilde A \tilde B}$}


\[
 {\tilde G}^H_{\tilde A \tilde B,\tilde D}K^{\tilde D}_{\alpha}+{\tilde G}^H_{\tilde R \tilde B}K^{\tilde R}_{\alpha,\tilde A}+{\tilde G}^H_{\tilde A \tilde R}K^{\tilde R}_{\alpha ,\tilde B}=0
\]

\begin{flushleft}
         $\;\;\;\;\;\;{\tilde A}\to A,\;\;\;{\tilde B} \to B$
\end{flushleft}

\begin{flushleft}
         $\rm {{\Roman 1}}\,.$ $\;\;\;\;{\tilde G}^H_{A B,D}K^{D}_{\alpha}+{\tilde G}^H_{A B,p}K^{p}_{\alpha}+{\tilde G}^H_{R B}K^{R}_{\alpha,A}+{\tilde G}^H_{AR }K^{R}_{\alpha,B}=0.$
\end{flushleft}

\begin{flushleft}
         $\;\;\;\;\;\;{\tilde A}\to p,\;\;\;{\tilde B} \to q$
\end{flushleft}

 \begin{flushleft}
         $\rm {{\Roman 2}}$\,. $\;\;\;\; {\tilde G}^H_{pq,D}K^{D}_{\alpha}+{\tilde G}^H_{pq,r}K^{r}_{\alpha}+{\tilde G}^H_{rq}K^{r}_{\alpha,p}+{\tilde G}^H_{pr }K^{R}_{\alpha,q}=0.$
\end{flushleft}

\begin{flushleft}
         $\;\;\;\;\;\;{\tilde A}\to p,\;\;\;{\tilde B} \to B$
\end{flushleft}

 \begin{flushleft}
         $\rm {{\Roman 3}}$\,. $\;\;\;\;{\tilde G}^H_{pB,D}K^{D}_{\alpha}+{\tilde G}^H_{pB,r}K^{r}_{\alpha}+{\tilde G}^H_{rB}K^{r}_{\alpha,p}+
{\tilde G}^H_{pR}K^{R}_{\alpha,B}=0.$
\end{flushleft}

\begin{flushleft}
         $\;\;\;\;\;\;{\tilde A}\to B,\;\;\;{\tilde B} \to p$
\end{flushleft}

\begin{flushleft}
         $\rm {{\Roman 4}}$\,. $\;\;\;\;{\tilde G}^H_{Bp,D}K^{D}_{\alpha}+{\tilde G}^H_{Bp,r}K^{r}_{\alpha}+{\tilde G}^H_{Rp}K^{R}_{\alpha,B}+{\tilde G}^H_{Br}K^{r}_{\alpha,p}=0.$
\end{flushleft}

$\rm{\Roman 4}=\rm{\Roman 3}$

\end{document}